\documentclass[english,nofootinbib, floatfix, superscriptaddress, prd, twocolumn,longbibliography]{revtex4-1}
\usepackage[T1]{fontenc}
\usepackage[latin9]{inputenc}

\usepackage{pdfpages}

\makeatletter
\AtBeginDocument{\let\LS@rot\@undefined}
\makeatother

\usepackage{graphicx,color}
\usepackage{subfigure}
\usepackage{amsmath,amsfonts,amssymb}
\usepackage{array}
\newcolumntype{P}[1]{>{\centering\arraybackslash}p{#1}}
\newcolumntype{M}[1]{>{\centering\arraybackslash}m{#1}}
\numberwithin{equation}{section}

\usepackage{hyperref}
\def\BCS{\textrm{BCS}}
\def\Eth{\textrm{Eth}}
\def\cl{\textrm{cl}}

\usepackage{comment}
\def\comment#1{}

\begin{document}

\title{Electrical conductivity and nuclear magnetic resonance relaxation rate of Eliashberg superconductors in the weak-coupling limit}

\author{Rufus Boyack}
\email{rufus.boyack@dartmouth.edu}
\affiliation{Department of Physics and Astronomy, Dartmouth College, Hanover, New Hampshire, 03755, USA}

\author{Sepideh Mirabi}
\affiliation{Department of Physics, Simon Fraser University, Burnaby, British Columbia, V5A 1S6, Canada}

\author{F. Marsiglio}
\affiliation{Department of Physics, University of Alberta, Edmonton, Alberta, T6G 2E1, Canada}

\begin{abstract}
Electrical conductivity is an important transport response in superconductors, enabling clear signatures of dynamical interactions to be observed. Of primary interest in this paper is to study characteristics of the electron-phonon interaction in weak-coupling Eliashberg theory (Eth), and to note the distinctions with Bardeen-Cooper-Schrieffer (BCS) theory. Recent analysis of weak-coupling Eth has shown that while there are modifications from the BCS results, certain dimensionless ratios are in agreement. Here we show that the conductivities in BCS theory and Eth fundamentally differ, with the latter having an imaginary gap component that damps a divergence. We focus on the dirty limit, and for both BCS theory and Eth we derive expressions for the low-frequency limit of the real conductivity. For Eth specifically, there are two limits to consider, depending on the relative size of the frequency and the imaginary part of the gap. In the case of identically zero frequency, we derive an analytical expression for the nuclear magnetic resonance relaxation rate. Our analysis of the conductivity complements the previous study of the Meissner response and provides a thorough understanding of weak-coupling Eth. 
\end{abstract}

\date{\today}
\maketitle

\section{Introduction}
The defining characteristics of superconductivity are~\cite{RickayzenBook, Scalapino1993}: (1) the Meissner effect (perfect diamagnetism) and (2) infinite DC electrical conductivity (perfect conductivity). The Meissner effect is a zero-frequency response as the wave vector approaches zero, whereas the zero-resistance state is determined from a zero-wave-vector response as the frequency approaches zero. Characterizing superconductivity in materials thus requires observation of flux expulsion (or flux trapping) in addition to the presence of zero DC resistivity. Theoretical calculations of conductivity are particularly challenging, in part because of the need to perform analytical continuation to real frequencies. 

The electromagnetic response of superconductors governed by Eliashberg theory (Eth)~\citep{Eliashberg1960, Eliashberg1961, Bardeen1973, Carbotte1990, Chubukov2020, Marsiglio2020} is of particular interest because it can provide a means to observe signatures of the dynamical electron-phonon interaction~\cite{Grimvall1976, Bennemann}. In previous papers~\cite{Marsiglio2018, Mirabi2020, Mirabi2020b} we investigated the critical temperature, the gap function on the real frequency axis, and the specific heat, all in the weak-coupling limit. These latter two papers thus encapsulate the Meissner response (a superconducting gap necessarily leading to a Meissner effect~\citep{Bardeen1957}) and the thermodynamic response. The present paper aims to provide an analysis of the conductivity in weak-coupling Eth, which will complete our understanding of points (1) and (2) mentioned above. 

There is a vast collection of articles studying the electrical conductivity within the Bardeen-Cooper-Schrieffer (BCS) and Eliashberg theories of superconductivity. In BCS theory, the conductivity has been studied in the clean~\cite{AGDBook} and dirty limits~\cite{Mattis1958}, and also for finite impurity-concentration strengths~\cite{Scharnberg1978, Leplae1983, Zimmerman1991, Chen1993}. In the case of Eth, different formalisms including Green's functions techniques~\cite{AGDBook} and quasiclassical methods~\cite{Rainer1995} have been utilised. The calculation of the conductivity for a frequency-dependent pairing function was first performed by Nam~\cite{Nam1967,Nam1967b} using the Green's function formalism. Shortly after Nam's work, Shaw and Swihart~\cite{Shaw1968} performed a similar strong-coupling analysis for Pb and Sn. The quasiclassical approach was utilised by Lee, Rainer, and Zimmermann~\cite{Lee1989} to directly compute conductivity on the real frequency axis. 

Bickers et al.~\cite{Bickers1990} computed the electrical conductivity on the imaginary frequency axis and used a Pad\'e approximant to perform the analytical continuation to the real axis. This method was also used by Nicol et al.~\cite{Nicol1991}, who extended the theory by including charge and spin fluctuations. Carbotte and collaborators~\cite{Akis1991, Akis1991c} applied the hybrid real and imaginary-frequency axes formulation~\cite{Marsiglio1988} of Eliashberg theory to the formalism developed by Lee et al., studying the conductivity at arbitrary temperatures and impurity concentrations. The conductivity of typical strong-coupling superconductors (Nb and Pb)~\cite{Marsiglio1991,Klein1994,Marsiglio1994} and  anisotropic superconductors~\cite{Jiang1996} has also been investigated. The conductivity sum rule~\citep{Marsiglio1995, Marsiglio1997, Chubukov2003, Marsiglio2008} and quasiparticle lifetimes~\citep{Kaplan1976,Kaplan1977,Marsiglio1997b} are also important topics. 

In the dirty limit, the electrical conductivity at zero frequency is proportional to the nuclear magnetic resonance (NMR) relaxation rate. In BCS theory this quantity has a logarithmic singularity~\citep{TinkhamBook, AlexandrovBook}, whereas in Eth the imaginary part of the gap ensures there is no divergence~\cite{Fibich1965, Fibich1965b}. Damping effects~\cite{Akis1991b,Choi1996} and anisotropic gap features~\cite{Statt1990} play an important role in the NMR relaxation rate. Disorder effects in NMR have been reported to be important~\cite{Choi1995}, however, other authors claim they are unimportant~\cite{Samokhin2006}. 

Here we complement our earlier works~\cite{Marsiglio2018, Mirabi2020, Mirabi2020b} on weak-coupling Eth by now considering the electrical conductivity. There has been recent renewed interest in understanding the weak-coupling limit of Eth~\cite{Yuzbashyan2022}, and since optical response is a clear way to observe dynamical features associated with the pairing interaction, theoretical predictions for weak-coupling response are needed. In addition, the NMR relaxation rate is studied in a myriad of superconductivity applications, with the absence of a peak in this signal being related to the absence of electron-phonon-driven superconductivity~\cite{Allen1991}. Consequently, it is crucial to understand the coherence peak in the NMR relaxation rate. We correct earlier work~\cite{Fibich1965,Fibich1965b} and provide a formula for the NMR relaxation rate in the dirty limit. The analysis is quite general, being valid for a system with a frequency-dependent (but momentum independent) complex gap. Other systems~\cite{Parker2008} with a similar gap structure will have an analogous formulation.

\section{Results}
\subsection{Theoretical analysis}
\label{sec:Theory}

\subsubsection{Review of weak-coupling Eliashberg theory}
\label{sec:WeakCoupling}

Before we begin our analysis of the electrical conductivity, here we provide a brief overview of weak-coupling Eth and its fundamental differences with BCS theory. More extensive analysis can be found in Refs.~\citep{Marsiglio2018, Mirabi2020, Mirabi2020b, Marsiglio2020}. In these two theories of superconductivity, the quantities of most interest are the pairing parameter $\Delta$ and the transition temperature $T_{c}$. In the most general context, $\Delta$ is a function of Matsubara frequencies $\omega_{m}$ (on the imaginary frequency axis) and momentum $\mathbf{k}$. For Eth, the quasiparticle residue $Z$ is also an important quantity, whereas in BCS theory $Z=1$. Let us consider only  $\Delta\left(\mathbf{k}, i\omega_{m}\right)$ and compare Eth and BCS theory.

Our schematic argument here follows Appendix 2.1 of Ref.~\citep{RickayzenBook}.
For a system of electrons interacting with phonons, the self-consistent equation for the pairing gap involves a momentum and frequency-dependent interaction $\lambda_{\mathbf{k}-\mathbf{k}^\prime}D\left(\mathbf{k}-\mathbf{k}^\prime, \omega_{m}-\omega^{\prime}_{m}\right)$. If the interaction is instantaneous, then $D$ is independent of $\omega_{m}-\omega^{\prime}_{m}$ and thus the gap is independent of frequency. The resulting self-consistent equation reproduces the standard~\cite{Bardeen1957} BCS gap equation. To evaluate this gap equation, the simplest assumption (see Sec.~5.3 in Ref.~\citep{RickayzenBook}) is to consider an interaction that has a cutoff in momentum space. On the other hand, if, instead, the interaction $\lambda D$ is independent of $\mathbf{k}-\mathbf{k}^{\prime}$, then the gap depends only on frequency. This is the usual Eth scenario. If one now attempts to use a model for $D\left(\omega_{m}-\omega^{\prime}_{m}\right)$ similar to that in BCS, namely a constant with a hard cutoff, but now in frequency space, then one obtains exactly the same equation as in BCS theory -- see page 418 of Ref.~\citep{RickayzenBook}. 

The above analogy has led to the belief~\citep{RickayzenBook} that, in the weak-coupling limit, the electron-phonon model with a hard cutoff in frequency space (Eth with a hard cutoff) is equivalent to the BCS case with a hard cutoff in momentum space. In this paper, however, weak-coupling Eth means the theory of electron-phonon superconductivity where the frequency dependence of the interaction $D$ is retained for all frequencies, but where the dimensionless electron-phonon coupling constant $\lambda_{\mathbf{k}-\mathbf{k}^\prime} \equiv\lambda\rightarrow0$. As shown in recent papers~\citep{Marsiglio2018, Mirabi2020, Mirabi2020b} by our group, in this limit there are quantitative differences between Eth and BCS theory. The important distinctions that concern us are in the transition temperature and the zero-temperature gap edge (to be defined more rigorously later on). The transition temperature in BCS theory (with $Z$ included~\citep{Marsiglio2018}) and weak-coupling Eth are given by~\citep{Marsiglio2018}:
\begin{align}
\label{eq:Tcs}
\frac{T_{c}^{\BCS}}{\omega_{E}} &= \frac{\exp\left[-\psi\left(1/2\right)\right]}{\left(2\pi\right)}\exp\left(-\frac{1+\lambda}{\lambda}\right), \\
\frac{T_{c}^{\Eth}}{\omega_{E}} &= \frac{1}{\sqrt{e}}\frac{T_{c}^{\BCS}}{\omega_{E}},
\end{align}
where $\psi$ is the digamma function, $\psi(1/2)\equiv-2\ln2-\gamma$, and $\gamma\approx 0.57722...$ is the Euler-Mascheroni constant~\cite{NIST2020}.

Similarly, for the zero-temperature gap edge~\citep{Mirabi2020}:
\begin{align}
\label{eq:GapEdges1}
\frac{\Delta_{0}^{\BCS}}{\omega_{E}} &= 2\exp\left(-\frac{1+\lambda}{\lambda}\right), \\
\frac{\Delta_{0}^{\Eth}}{\omega_{E}} &= \frac{1}{\sqrt{e}}\frac{\Delta_{0}^{\BCS}}{\omega_{E}}.
\label{eq:GapEdges2}
\end{align}
Here, $\omega_{E}$ is the Einstein frequency in an Einstein model for the electron-phonon interaction. The prefactor $1/\sqrt{e}$ causes the weak-coupling Eth expressions to differ from their BCS counterparts. Interestingly, the dimensionless ratio $\Delta_{0}/T_{c}$ is the same for both theories. This was already noted in Ref.~\citep{Mitrovic1984}. In the present paper, our interest is in studying transport quantities where the explicit distinction between weak-coupling Eth and BCS theory does matter. As noted earlier, electrical conductivity is a natural transport response to consider because it depends on the frequency of the external perturbation. Therefore, a dynamical theory like Eth would be expected to behave differently from BCS theory. As we shall find, there is indeed a quantifiable (and presumably measurable) difference in the electrical conductivities of these theories. In the next sections we present our theoretical and numerical calculations.

\subsubsection{Electrical conductivity in the dirty limit}
\label{sec:DirtyLimit}

Our numerical and theoretical analyses of electrical conductivity are based on the formulas given in Refs.~\cite{Lee1989, Marsiglio1991, Akis1991, Klein1994, Marsiglio1994, Marsiglio1995, Tajik2019}. The derivation of the conductivity is similar to that of the phonon self energy, which is derived in Ref.~\citep{Marsiglio1992}. An alternative derivation of the conductivity, based on the standard method~\cite{AGDBook}, can be found in Ref.~\citep{Marsiglio1991}. This latter reference expresses the conductivity in a different manner than the previous references by isolating the normal-state contribution. 

Let $\nu$ denote the external frequency and $1/\tau$ the impurity scattering rate, which we assume to be elastic. Then, $\sigma(\nu)$ is given by~\cite{Klein1994}:
\begin{widetext}
\begin{align}
\label{eq:KleinSigma}
\sigma\left(\nu\right) & = \frac{\omega_{p}^{2}}{8\pi\nu}\left\{ \int_{0}^{\infty}\tanh\left(\frac{\omega}{2T}\right)
\left[\frac{1-N\left(\omega\right)N\left(\omega+\nu\right)-M\left(\omega\right)M\left(\omega+\nu\right)}{-i\epsilon\left(\omega\right)-i\epsilon\left(\omega+\nu\right)
+1/\tau}\right]d\omega\right.\nonumber \\
 & \quad  +\int_{0}^{\infty}\tanh\left(\frac{\omega+\nu}{2T}\right)
 \left[\frac{1-N^{*}\left(\omega\right)N^{*}
 \left(\omega+\nu\right)-M^{*}\left(\omega\right)M^{*}
 \left(\omega+\nu\right)}{-i\epsilon^{*}\left(\omega\right)-i\epsilon^{*}\left(\omega+\nu\right)-1/\tau}\right]d\omega \nonumber \\
 & \quad +\int_{0}^{\infty}\left[\tanh\left(\frac{\omega+\nu}{2T}\right)-\tanh\left(\frac{\omega}{2T}\right)\right]\left[\frac{1+N^{*}\left(\omega\right)N\left(\omega+\nu\right)+M^{*}\left(\omega\right)M\left(\omega+\nu\right)}{i\epsilon^{*}\left(\omega\right)-i\epsilon\left(\omega+\nu\right)+1/\tau}\right]d\omega \nonumber \\
 & \quad +\int_{-\nu}^{0}\tanh\left(\frac{\omega+\nu}{2T}\right)\left[\frac{1-N^{*}\left(\omega\right)N^{*}\left(\omega+\nu\right)-M^{*}\left(\omega\right)M^{*}\left(\omega+\nu\right)}{-i\epsilon^{*}\left(\omega\right)-i\epsilon^{*}\left(\omega+\nu\right)-1/\tau}\right.\nonumber \\
 & \quad  \left.\left. +\frac{1+N^{*}\left(\omega\right)N\left(\omega+\nu\right)+M^{*}\left(\omega\right)M\left(\omega+\nu\right)}{i\epsilon^{*}\left(\omega\right)-i\epsilon\left(\omega+\nu\right)+1/\tau}\right]d\omega\right\} .
\end{align}
\end{widetext}
Here, $\omega_{p}=\sqrt{4\pi ne^2/m}$ is the plasma frequency, where $m$ is the electron mass, $e$ is the elementary charge, and $n$ is the number density. We use Natural units: $\hbar=c=k_{B}=1$. The functions $N$ and $M$ are defined by
\begin{align}
\epsilon\left(\omega\right) &= \sqrt{\omega^{2}Z^{2}\left(\omega+i\delta\right) - \phi^{2}\left(\omega+i\delta\right)}. \\
N\left(\omega\right)	&= \frac{\omega Z\left(\omega+i\delta\right)}{\epsilon\left(\omega\right)}. \\
M\left(\omega\right)	&= \frac{\phi\left(\omega+i\delta\right)}{\epsilon\left(\omega\right)}.
\end{align}
Note that $\phi(\omega+i\delta)\equiv\Delta(\omega+i\delta)Z(\omega+i\delta)$. The sign of the square root is chosen~\cite{Allen1983,Marsiglio1988} such that the imaginary part is positive. In Eq.~\eqref{eq:KleinSigma}, the $\phi$ and $Z$ that appear are those determined in the clean limit. The Eth gap equations for $\Delta\left(\omega+i\delta\right)$ and $Z\left(\omega+i\delta\right)$ in the clean case are given in Ref.~\citep{Marsiglio1991}; we also present these gap equations in Supplementary Note 1. The effects from impurities are completely encapsulated within the $1/\tau$ terms. As mentioned earlier, in Refs.~\cite{Tajik2019,Marsiglio1995} the $1/\tau$ term is not written explicitly, rather, the $\phi$ and $Z$ parameters used are those for a  system with impurities~\cite{Marsiglio1992}:
\begin{align}
\phi(\omega) &= \phi_{\cl}(\omega)+\frac{i}{2\tau}\frac{\Delta(\omega)}{\sqrt{\omega^2-\Delta^2(\omega)}}, \\
Z\left(\omega\right) &= Z_{\cl}\left(\omega\right)+\frac{i}{2\tau}\frac{1}{\sqrt{\omega^2-\Delta^2(\omega)}}.
\end{align}
For these expressions, the real part of $\epsilon$ must have the same sign as $\omega$~\cite{Allen1983,Marsiglio1992}. If one inserts the definition of $\phi$ in terms of $\Delta$ and $Z$, then one finds that $\Delta(\omega)=\Delta_{\cl}(\omega)$ and $\epsilon(\omega)=Z_{\cl}(\omega)\sqrt{\omega^2-\Delta^2_{\cl}(\omega)}+i/(2\tau)\equiv\epsilon_{\cl}(\omega)+i/(2\tau)$. Thus, the electrical conductivity can be studied at arbitrary impurity strengths by inserting the clean Eliashberg parameters into Eq.~\eqref{eq:KleinSigma}, with the impurity scattering dependence completely encapsulated in the $1/\tau$ terms. That $\Delta$ is unaffected by impurities is in accord with Anderson's observation~\citep{Anderson1959} that the transition temperature is independent of impurity concentration for weak impurity scattering and for impurities that preserve time-reversal symmetry. A simple illustration of this observation can be found in Ref.~\citep{Parks2}. Paramagnetic impurities thus play a role analogous to a magnetic field~\cite{Maki1965}.  

The contribution to elastic scattering arises from $\tau$; within the context of Eth, there is also a contribution to inelastic scattering arising from the electron-phonon interaction. Indeed, in Eth the total scattering rate is given by~\citep{Marsiglio1991}: $\widetilde{\tau}^{-1}=\tau^{-1}+2\pi\int_{0}^{\infty}\alpha^2F(\omega)\coth[\omega/(2T)]d\omega$, where $\alpha^2F(\omega)$ is the electron-phonon spectral function; here we shall consider an Einstein model with $\alpha^2 F(\omega)=\frac{1}{2}\lambda\omega_{E}\delta(\omega-\omega_{E})$, where $\lambda$ is the dimensionless coupling constant. Note that, the two contributions to scattering are added in accord with Matthiessen's rule~\citep{Matthiessen1864,ZimanBook} (the total scattering rate is the sum of the individual scattering rates; equivalently, scattering times ``add in parallel''). Interestingly, the electron-phonon interaction does not modify the DC electrical conductivity (in the clean limit)~\citep{Nakajima1963,Prange1964,Grimvall1976}. For more discussion on the normal-state electrical conductivity, see Refs.~\citep{Shulga1991,Marsiglio1995}. 

It is important to realize that the introduction of $\nu$ and $\tau$ means that the optical conductivity depends on two external energy scales ($\hbar\nu$ and $\hbar/\tau$). In the case of the zero-temperature gap edge~\cite{Mirabi2020} and the specific heat~\cite{Mirabi2020b}, it was possible to obtain dimensionless ratios that have the same universal value in both weak-coupling Eth and BCS theory. However, there are no universal ratios involving the conductivity. Indeed, the units for optical conductivity (in three spatial dimensions) are $s^{-1}$, and one natural scale is the DC conductivity for a system with finite impurity scattering~\cite{ZimanBook}: $\sigma_{0}\left(\tau\right)\equiv ne^2\tau/m$.  

For a given coupling constant and temperature, the conductivity is a function of the dimensionless quantities $\nu/E_{0}$ and $1/(E_{0}\tau)$, where $E_{0}$ is an unspecified energy scale. In BCS theory, it is natural to use the zero-temperature gap for $E_{0}$: $E_{0}=\Delta^{\BCS}_{0}$, whereas in Eth it is natural to use the Einstein frequency: $E_{0}=\omega_{E}$. Nevertheless, for both theories one could define $E_{0}$ as the zero-temperature gap edge. However, it has been shown~\cite{Marsiglio2018,Mirabi2020} that the gap edge differs between BCS theory and weak-coupling Eth. Hence, it is not possible to use equivalent values for the independent variables in the conductivities of BCS and Eth; said in a different way, equivalent independent variables would require different respective interaction strengths. Thus, the optical response fundamentally differs between BCS and Eth, and any signatures in conductivity should then indicate an appropriate energy scale that can be related to the interactions (Einstein frequency etc). To facilitate a meaningful comparison between BCS and Eth, we can attempt to mitigate the dependence of $\sigma$ on $\nu$ and $\tau$. First let us consider $\tau$.

If $\ell$ denotes the mean-free path and $\xi$ the coherence length, then the two natural cases to consider are~\citep{Klein1994}: (1) the clean limit, $\ell/\xi\gg1$, and (2) the dirty (or local) limit, $\ell/\xi\ll1$. In the former case there are two subcases (a) the Pippard limit (clean type-I superconductor) and (b) the London limit (clean type-II superconductor). In the dirty-limit case, since the mean-free path is very small compared to the coherence length, quasiparticle scattering happens on a very short time scale, which means that $\tau\ll1/E_{0}$. As discussed in Ref.~\cite{Nam1967}, the normalized conductivity is equivalent in the local and anomalous limits. Thus, for the strong-coupling superconductors Nb (local limit) and Pb (anomalous limit), the local limit is a valid approximation to consider. Our theoretical analysis from here on will thus focus on $E_{0}\tau\ll1$. 

Let us define $\widetilde{\sigma}_{1}(\nu)\equiv\mathrm{Re}[\sigma\left(\nu\right)]/\sigma_{0}$. Taking the dirty-limit in Eq.~\eqref{eq:KleinSigma} leads to:
\begin{align}
\label{eq:DirtySigma}
\widetilde{\sigma}_{1}(\nu) & = \frac{1}{\nu}\int_{0}^{\infty}\biggl\{
\biggl[\mathrm{Re}N\left(\omega\right)\mathrm{Re}N\left(\omega+\nu\right) \nonumber\\
&\hspace{1.47cm} +\mathrm{Re}M\left(\omega\right)\mathrm{Re}M\left(\omega+\nu\right)\biggr] \nonumber\\
&\quad \times \left[\tanh\left(\frac{\omega+\nu}{2T}\right)-\tanh\left(\frac{\omega}{2T}\right)\right]\biggr\} d\omega \nonumber \\
 & \quad +\frac{1}{\nu}\int_{-\nu}^{0}\biggl\{
 \biggl[\mathrm{Re}N\left(\omega\right)\mathrm{Re}N\left(\omega+\nu\right) \nonumber\\
&\hspace{1.72cm} +\mathrm{Re}M\left(\omega\right)\mathrm{Re}M\left(\omega+\nu\right)\biggr] \nonumber\\
&\quad \times \tanh\left(\frac{\omega+\nu}{2T}\right)\biggr\} d\omega.
\end{align}

In the case where the gap is frequency independent, Eq.~\eqref{eq:DirtySigma} reduces to the Mattis-Bardeen~\cite{Mattis1958} result:
\begin{align}
\label{eq:MattisBardeen}
\widetilde{\sigma}^{\BCS}_{1}(\nu)& = \frac{1}{\nu}\int_{\Delta}^{\infty}\biggl\{
\frac{\omega\left(\omega+\nu\right)+\Delta^{2}}{\sqrt{\omega^{2}-\Delta^{2}}
\sqrt{\left(\omega+\nu\right)^{2}-\Delta^{2}}} \nonumber\\
&\quad \times\left[\tanh\left(\frac{\omega+\nu}{2T}\right)-\tanh\left(\frac{\omega}{2T}\right)\right]\biggr\} d\omega \nonumber \\
& \quad -\frac{1}{\nu}\int_{\Delta-\nu}^{-\Delta}\biggl\{
\frac{\omega\left(\omega+\nu\right)+\Delta^{2}}{\left|\sqrt{\omega^{2}-\Delta^{2}}\right|\sqrt{\left(\omega+\nu\right)^{2}-\Delta^{2}}} \nonumber\\
&\quad \times\Theta\left(\nu-2\Delta\right)\tanh\left(\frac{\omega+\nu}{2T}\right)\biggr\} d\omega.
\end{align}
Note that, the term $\sqrt{\omega^2-\Delta^2}$ in the second line of Eq.~\eqref{eq:MattisBardeen} has a negative sign because $\omega$ is negative in the particular region of integration. We have explicitly accounted for this negative sign in front of the second term in Eq.~\eqref{eq:MattisBardeen} and written the square root within an absolute value. In other literature~\citep{Bennemann} the implicit negative sign in the square root is retained, and so the formulas are equivalent. For BCS theory, Eq.~\eqref{eq:MattisBardeen} can be evaluated analytically at $T=0$; see the Methods subsection titled `A summary of results for the BCS electrical conductivity'. At $T=0$, one finds that the real part of the response is non-zero only for $\nu\geq2\Delta$, for all impurity strengths. In Ref.~\citep{Zimmerman1991}, this frequency gap in the zero-temperature conductivity was observed in the BCS response. More recently, researchers have investigated additional spectral-weight contributions in the low-frequency response of strongly disordered superconductors~\citep{Seibold2017, Pracht2017} and have generalized the Mattis-Bardeen formula to include an energy-dependent density of states.

The expressions above indicate that, in the dirty limit, the real part of the conductivity depends on the elastic impurity scattering in proportion to $\sigma_{0}\sim\tau$. Intuitively this can be understood~\cite{RickayzenBook} from the fact that (elastic) impurities can scatter quasiparticles and thus they can reduce the current carried by quasiparticles. Further discussion on the role of impurities in optical response can be found in Ref.~\citep{Maslov2017}, and the effects of spatial randomness in electron-phonon models are discussed in Ref.~\citep{Guo2022}.

Random impurities that scatter elastically do not affect the velocity of Cooper pairs, nor can they destroy Cooper pairs and create quasiparticles. Moreover, in the dirty limit, where $\ell\ll\xi$, impurities that scatter elastically do not affect the wave function. Inelastic scattering from the electron-phonon interaction, however, does provide a mechanism for destroying Cooper-pairs. We shall discuss the consequences of this physics in more detail in the next section where we shall discover that inelastic scattering ensures that Eq.~\eqref{eq:DirtySigma} does not diverge in the $\nu\rightarrow0$ limit. This limit shall be carefully defined shortly. 

\subsubsection{Low-frequency analysis}
\label{sec:DCLimit} 

In the previous section we obtained a general expression~\eqref{eq:DirtySigma} for the real part of the electrical conductivity in the dirty limit. The impurity scattering rate no longer appears in the integrand, and the only external variable remaining (not including the coupling constant and the temperature) is the frequency $\nu$.  

In BCS theory, the low-frequency limit can be defined as $\nu/\Delta^{\BCS}_{0}\rightarrow0$. As can be proved~\cite{Parks1, TinkhamBook, Bennemann} using Eq.~\eqref{eq:MattisBardeen}, the conductivity diverges logarithmically in this limit. The mathematical form of this logarithmic singularity was obtained by Cullen and Ferrell~\citep{Cullen1966} in their study of attenuation of transverse ultrasound in superconductors, and also by Rogovin and Scalapino~\citep{Rogovin1974} in the context of fluctuation phenomena in tunnel junctions. In the Methods subsection titled `Low-frequency limit of the Mattis-Bardeen formula', we present our own derivation of the complete low-frequency response, in order to address all temperatures in the range $0\leq T\leq T_{c}$. 

For Eth, the low-frequency limit is more subtle. Indeed, as noted by earlier investigators~\citep{Fibich1965, Fibich1965b, Parks1, Allen1991}, due to absorption of thermal phonons there is a broadening of the quasiparticle lifetime and this precludes the appearance of a logarithmic divergence in the conductivity. Indeed, in Eth the gap is a complex function, and so, provided the imaginary part of the gap is sufficiently large, there is no logarithmic singularity in the low-frequency conductivity because the integrand in Eq.~\eqref{eq:DirtySigma} is never singular. Hence, we must define the low-frequency limit of Eq.~\eqref{eq:DirtySigma} in a careful manner. 

Let $\Delta_{1}(T)$ denote the gap edge and $\Delta_{2}(T)$ the corresponding imaginary part of the gap, defined by~\cite{Fibich1965}:
\begin{align}
\label{eq:Delta1Def}
\textrm{Re}\left[\Delta(\omega=\Delta_1,T)\right] &= \Delta_{1}(T).  \\
\textrm{Im}\left[\Delta(\omega=\Delta_1,T)\right] &= \Delta_{2}(T).
\label{eq:Delta2Def}
\end{align} 
The definition of $\Delta_{1}$ arises from solving for the singularity of the quasiparticle density of states, which is proportional to $\textrm{Re}\left[\omega/\sqrt{\omega^2-\Delta^{2}\left(\omega\right)}\right]$. In BCS theory the gap parameter is purely real and has no frequency dependence, and so the gap edge trivially reduces to the temperature-dependent gap parameter, and also $\Delta^{\BCS}_{2}=0$. We shall let $\Delta^{\Eth}_{1}$ and $\Delta^{\BCS}_{1}$ denote the (finite-temperature) gap edges in Eliashberg theory and BCS theory, respectively. Similarly, we let $\Delta^{\Eth}_{0}$ and $\Delta^{\BCS}_{0}$ denote the zero-temperature gap edges in the respective theories. For Eth response,  
there are two pertinent limits to consider: (1) $\Delta^{\Eth}_{2}\ll\nu\ll\Delta^{\Eth}_{1}$ and (2) $\nu\ll\Delta^{\Eth}_{2}\ll\Delta^{\Eth}_{1}$. The former limit is a BCS-like low-frequency limit where the imaginary part of the gap is small on the scale of $\nu$, which will mean that thermal phonon broadening is not important. The second limit is unique to Eth and corresponds to the case where the imaginary part of the gap is large enough to allow one to directly set $\nu=0$ in Eq.~\eqref{eq:DirtySigma}. 

For case (2), this DC response is also known as the NMR relaxation rate, and it is given by~\cite{Parks1} 
\begin{align}
\label{eq:EthNMR}
\frac{R_{s}}{R_{n}}&= 2\int_{0}^{\infty}\left\{ \left[\mathrm{Re}N\left(\omega\right)\right]^{2}
+\left[\mathrm{Re}M\left(\omega\right)\right]^{2}\right\} \left(-\frac{\partial f(\omega)}{\partial\omega}\right)d\omega. 
\end{align}
Here $f(x)=[\exp(x/T)+1]^{-1}$ is the Fermi-Dirac distribution function, and we define $f^\prime(x)\equiv\partial f(x)/\partial x$. The symbol for the NMR relaxation rate is $1/T_{1}$, and the dimensionless quantity $R$ (proportional to $\sigma(\nu=0)$) is defined by $R\equiv1/(T_{1}T)$. In the normal-state, the proportionality of $1/T_{1}$ to $T$ is known as Korringa's law~\citep{Korringa1950}. As shown in the Methods subsection titled `Low-frequency limit of the Mattis-Bardeen formula', the low-frequency limit of the conductivity in BCS theory is given by
\begin{align}
\label{eq:BCSDC}
\widetilde{\sigma}^{\BCS}_{1}\left(\nu\right) &=2f\left(\Delta^{\BCS}_{1}\right)-2\Delta^{\BCS}_{1} f^{\prime}\left(\Delta^{\BCS}_{1}\right)\log\left(\frac{8\Delta^{\BCS}_{1}}{\nu}\right)  \nonumber\\
&\quad+4\Delta^{\BCS}_{1}\int_{1}^{\infty}\frac{f^{\prime}\left(\Delta^{\BCS}_{1}\right)-f^{\prime}\left(x\Delta^{\BCS}_{1}\right)}{x^{2}-1}dx, \nonumber\\ 
&\quad\textrm{where}\ \nu\ll\Delta^{\BCS}_{1}.
\end{align}
For Eth, if $\Delta^{\Eth}_{2}\ll\nu$, then the analysis is similar to BCS theory. Indeed, the main contribution to the integral in Eq.~\eqref{eq:DirtySigma} occurs when $\omega\approx\Delta^{\Eth}_{1}$ (defined in Eq.~\eqref{eq:Delta1Def}). 
Thus, we can let $\Delta\approx\Delta^{\Eth}_{1}+i\Delta^{\Eth}_{2}\approx\Delta^{\Eth}_{1}$ and follow the BCS analysis to obtain:
\begin{align}
\label{eq:EthDC1}
\widetilde{\sigma}^{\Eth}_{1}\left(\nu\right) &= 2f\left(\Delta^{\Eth}_{1}\right) -2\Delta^{\Eth}_{1}f^{\prime}\left(\Delta^{\Eth}_{1}\right)\log\left(\frac{8\Delta^{\Eth}_{1}}{\nu}\right) \nonumber\\ &\quad +4\Delta^{\Eth}_{1}\int_{1}^{\infty}\frac{f^{\prime}\left(\Delta^{\Eth}_{1}\right)-f^{\prime}\left(x\Delta^{\Eth}_{1}\right)}{x^{2}-1}dx, \nonumber\\
&\quad \textrm{where}\ |\Delta^{\Eth}_{2}|\ll\nu\ll\Delta^{\Eth}_{1}.
\end{align}
In the case where $\nu=0$, the analysis is more involved. In the Methods subsection titled `Low-frequency limit for dirty Eliashberg superconductors', it is shown that the result is
\begin{align}
\label{eq:EthDC2}
\left(\frac{R_{s}}{R_{n}}\right)^{\Eth} &=2f\left(\Delta^{\Eth}_{1}\right) -2\Delta^{\Eth}_{1}f^{\prime}\left(\Delta^{\Eth}_{1}\right)\log\left(\frac{\sqrt{8}\Delta^{\Eth}_{1}}{\left|\Delta^{\Eth}_{2}\right|}\right) \nonumber\\ &\quad +4\Delta^{\Eth}_{1}\int_{1}^{\infty}\frac{f^{\prime}\left(\Delta^{\Eth}_{1}\right)-f^{\prime}\left(x\Delta^{\Eth}_{1}\right)}{x^{2}-1}dx,\nonumber\\
&\quad \textrm{where}\ |\Delta^{\Eth}_{2}|\ll\Delta^{\Eth}_{1}.
\end{align}

The BCS conductivity in Eq.~\eqref{eq:BCSDC} is a function of the BCS gap $\Delta^{\BCS}_{1}$. Likewise, the Eth conductivity in Eq.~\eqref{eq:EthDC1} is a function of the Eth gap $\Delta^{\Eth}_{1}$. As shown in Refs.~\citep{Marsiglio2018, Mirabi2020}, in the weak-coupling limit these two gaps can be related to one another. Indeed, Eqs.~\eqref{eq:GapEdges1}-\eqref{eq:GapEdges2} show that the zero-temperature gap edges are related by $\Delta^{\Eth}_{0}=\Delta^{\BCS}_{0}/\sqrt{e}$. Furthermore, from Eqs.~\eqref{eq:Tcs}-\eqref{eq:GapEdges2}, it follows that $\Delta^{\Eth}_{1}/T^{\Eth}_{c} =\Delta^{\BCS}_{1}/T^{\BCS}_{c}$. These relations mean that we can relate Eqs.~\eqref{eq:BCSDC} and \eqref{eq:EthDC1} to each other. Importantly, the reason why conductivities with different arguments can be related is precisely because, in the weak-coupling limit, the two gap edges and the two transition temperatures are related to each other. In Eq.~\eqref{eq:EthDC1}, we can normalize the gap to its zero-temperature value and the temperature to the corresponding $T_{c}$, defining $t\equiv T/T_{c}$ as the reduced temperature, and then take the weak-coupling limit:
\begin{align}
\frac{\Delta_{1}^{\Eth}}{T}	&=	\left(\frac{\Delta_{1}}{\Delta_{0}}\right)^{\Eth}\left(\frac{\Delta_{0}}{T_{c}}\right)^{\Eth}\frac{1}{t}\nonumber\\
&\rightarrow\left(\frac{\Delta_{1}}{\Delta_{0}}\right)^{\BCS}\left(\frac{\Delta_{0}}{T_{c}}\right)^{\BCS}\frac{1}{t}, \textrm{\ as\ } \lambda\rightarrow0.
\label{eq:Ratio1}
\end{align}
Because $\nu$ does not depend on $\lambda$, once we express $\nu$ in units of either the gap edge or the transition temperature, there will be weak-coupling corrections present. Indeed, 
\begin{align}
\frac{\Delta_{1}^{\Eth}}{\nu} &= \left(\frac{\Delta_{1}}{\Delta_{0}}\right)^{\Eth}\frac{\Delta_{0}^{\Eth}}{\nu} \nonumber\\
&\rightarrow\left(\frac{\Delta_{1}}{\Delta_{0}}\right)^{\BCS}\frac{\Delta_{0}^{\BCS}}{\nu}\frac{1}{\sqrt{e}}, \textrm{\ as\ } \lambda\rightarrow0.
\label{eq:Ratio2}
\end{align}
Using Eqs.~\eqref{eq:Ratio1}-\eqref{eq:Ratio2}, we can express the Eth conductivity \eqref{eq:EthDC1} in terms of the BCS conductivity \eqref{eq:BCSDC} as:
\begin{align}
\sigma^{\Eth}\left(\nu;\Delta_{1}^{\Eth}(t),t\right) &= \sigma^{\BCS}\left(\nu;\Delta^{\BCS}(t),t\right) \nonumber\\ &\quad +\Delta^{\BCS}(t)f^{\prime}\left(\Delta^{\BCS}(t)\right).
\label{eq:EthBCSSigmaComp}
\end{align}
This result illustrates that the electrical conductivity is not universal; there is a distinction between the conductivities in BCS and Eth. Indeed, the frequency $\nu$ in the arguments of the logarithms in Eqs.~\eqref{eq:BCSDC} and \eqref{eq:EthDC1} cannot be expressed as a universal dimensionless form. 

Fibich derived~\citep{Fibich1965,Fibich1965b} an expression for the NMR relaxation rate in Eth and he found a logarithmic dependence on the ratio $\Delta^{\Eth}_{1}/|\Delta^{\Eth}_{2}|$, where $\Delta^{\Eth}_{1}$ and $\Delta^{\Eth}_{2}$ are defined in Eqs.~\eqref{eq:Delta1Def}-\eqref{eq:Delta2Def}. However, we have performed the analysis independently from Fibich, and we find that our expression in Eq.~\eqref{eq:EthDC2} disagrees with Fibich's result in Ref.~\citep{Fibich1965b}. There was some skepticism~\citep{Allen1991} about the applicability of Fibich's formula in certain cases, noting that it was suitable provided an unrealistically large value of the Debye frequency was used. In the `Electrical conductivity' subsection of the Numerical results section, we present numerical calculations of the NMR relaxation rate using Eq.~\eqref{eq:EthNMR} and we find good agreement with our theoretical result Eq.~\eqref{eq:EthDC2}. 

The NMR relaxation rate is not easily defined in BCS theory because there is no damping term in the integrand of Eq.~\eqref{eq:MattisBardeen}. Hebel and Slichter~\cite{Hebel1957, Hebel1959} noted that the Zeeman energy would provide a way to prevent the divergence, but if this is too small then an artifical lifetime effect would need to be incorporated; this is tantamount to the introduction of a small cutoff in the lower limit of integration~\citep{Allen1991}. Eliashberg theory naturally provides a lifetime broadening effect, as noted by Fibich~\citep{Fibich1965}. Gap anisotropy~\citep{Parks1, Statt1990} is also a contributing factor to damping the singularity that would be present in the NMR relaxation for a BCS scenario.

\subsubsection{Weak-coupling analysis of \texorpdfstring{$\Delta^{\Eth}_{2}$}{}}
\label{Sec:Delta2}

In Eq.~\eqref{eq:EthDC2}, we have obtained an analytical expression for the Eth NMR relaxation rate. This expression depends on both $\Delta^{\Eth}_{1}(T)$ and $\Delta^{\Eth}_{2}(T)$, which need to be determined numerically from the Eth gap equations, and so the result is not yet in the form of a function that explicitly depends solely on temperature. In the weak-coupling limit, the normalized gap edge $\Delta^{\Eth}_{1}(T)/\Delta^{\Eth}_{1}(T=0)$ tends to its BCS value, which can be approximated by a simple interpolation function~\citep{RickayzenBook}. Having a closed-form formula for the NMR relaxation rate would facilitate easier computation and comparison with experimental results. 

There is a dearth of literature on calculations of $\Delta^{\Eth}_{2}$. Fibich~\citep{Fibich1965} performed some calculations for the metal Al. Scalapino and Wu~\citep{Scalapino1966} derived a general self-consistent equation for $\Delta^{\Eth}_{2}$ by using the Eliashberg gap equations. They focused on systems with a Debye spectrum for the electron-phonon coupling, and hence low frequencies were most relevant in their analysis. In Fig.~9 of Ref.~\cite{Maclaughlin1976} a comparison between the predictions of Fibich and Scalapino and Wu is provided. In addition, Refs.~\cite{Williamson1973, Kaplan1976, Kaplan1977} provided some comparison with experiment. However, the analysis of $\Delta^{\Eth}_{2}$ for an electron-phonon coupling with an Einstein spectrum, which is the focus of this work, does not appear to have been undertaken previously. Here, our aim is to address this problem and obtain a simple approximation for $\Delta^{\Eth}_{2}(T)$.  

As shown in Supplementary Note 1, the self-consistent equation for $\Delta^{\Eth}_{2}$ is given by
\begin{widetext}
\begin{align}
\label{eq:Delta2}
&Z^{\Eth}_{1}\Delta^{\Eth}_{2} =  -\frac{\pi}{2}\lambda\omega_{E} \textrm{Re}\left\{ N\left(\omega_{E}\right)\left[\frac{\Delta^{\Eth}_{1}+\omega_{E}-\Delta\left(\Delta^{\Eth}_{1}+\omega_{E}\right)}{\sqrt{\left(\Delta^{\Eth}_{1}+\omega_{E}\right)^{2}-\Delta^{2}\left(\Delta^{\Eth}_{1}+\omega_{E}\right)}}+\frac{\omega_{E}-\Delta^{\Eth}_{1}+\Delta\left(\omega_{E}-\Delta^{\Eth}_{1}\right)}{\sqrt{\left(\omega_{E}-\Delta^{\Eth}_{1}\right)^{2}-\Delta^{2}\left(\omega_{E}-\Delta^{\Eth}_{1}\right)}}\right]\right.\nonumber \\
 &\quad +\left.f\left(\omega_{E}+\Delta^{\Eth}_{1}\right)\frac{\Delta^{\Eth}_{1}+\omega_{E}-\Delta\left(\Delta^{\Eth}_{1}+\omega_{E}\right)}{\sqrt{\left(\Delta^{\Eth}_{1}+\omega_{E}\right)^{2}-\Delta^{2}\left(\Delta^{\Eth}_{1}+\omega_{E}\right)}}+f\left(\omega_{E}-\Delta^{\Eth}_{1}\right)\frac{\omega_{E}-\Delta^{\Eth}_{1}+\Delta\left(\omega_{E}-\Delta^{\Eth}_{1}\right)}{\sqrt{\left(\omega_{E}-\Delta^{\Eth}_{1}\right)^{2}-\Delta^{2}\left(\omega_{E}-\Delta^{\Eth}_{1}\right)}}\right\} .
\end{align}
\end{widetext}
While $\Delta^{\Eth}_{2}$ does not explicitly appear on the right-hand side of this equation, it must be kept in mind that it implicitly appears. Indeed, in Eq.~\eqref{eq:Delta2} there are terms of the form $\Delta\left(\omega_{E}\pm\Delta^{\Eth}_{1}\right)$; as shown in the Eth gap equations in the Supplemental material, $\Delta\left(\omega\right)$ depends on $\Delta\left(\omega\pm\omega_{E}\right)$. Thus, if we set $\omega=\omega_{E}\pm\Delta^{\Eth}_{1}$ in the aforementioned equations, then we shall find terms involving $\Delta(\Delta^{\Eth}_{1})\equiv\Delta^{\Eth}_{1}+i\Delta^{\Eth}_{2}$. As a result, the terms $\Delta\left(\omega_{E}\pm\Delta^{\Eth}_{1}\right)$ on the right-hand side of Eq.~\eqref{eq:Delta2} do in fact depend on $\Delta^{\Eth}_{2}$ via the Eth gap equations. Hence, Eq.~\eqref{eq:Delta2} is not an explicit expression for $\Delta^{\Eth}_{2}$. Incidentally, note that if we insert the Einstein form for the electron-phonon interaction into Eq.~(9) of Ref.~\citep{Scalapino1966}, then we reproduce the result in Eq.~\eqref{eq:Delta2}. 

In the weak-coupling limit~\cite{Mirabi2020b}, $Z^{\Eth}_{1}\approx1+\lambda$. Since $T_{c}\ll\omega_{E}$, in Eq.~\eqref{eq:Delta2} we can replace the Fermi-Dirac and Bose-Einstein distribution functions by $e^{-\frac{\omega_{E}}{T_{c}t}}$. 
In order to derive a closed-form expression for $\Delta^{\Eth}_{2}$, we need to determine $\Delta(\Delta^{\Eth}_{1}\pm\omega_{E})$. However, the gap drastically changes in the vicinity of $\omega_{E}$, with $\Delta(\omega_{E})$ being singular and $\Delta(\Delta^{\Eth}_{1}\pm\omega_{E})/\omega_{E}\sim\mathcal{O}(1)$. In Supplementary Note 1, we present further discussion of our attempt to determine $\Delta^{\Eth}_{2}$ analytically by applying the approximate real-axis expressions in Ref.~\citep{Mirabi2020}. In the next section we shall numerically determine $\Delta^{\Eth}_{2}$ and use this result to compute the NMR relaxation rate in Eq.~\eqref{eq:EthDC2} and compare with the numerical result in Eq.~\eqref{eq:EthNMR}.

\subsection{Numerical results}
\label{sec:NumericalResults}

\subsubsection{Electrical conductivity}
\label{sec:OpticalResponse}

\begin{figure}[h]
\centering\includegraphics[width=7.5cm,height=11cm]{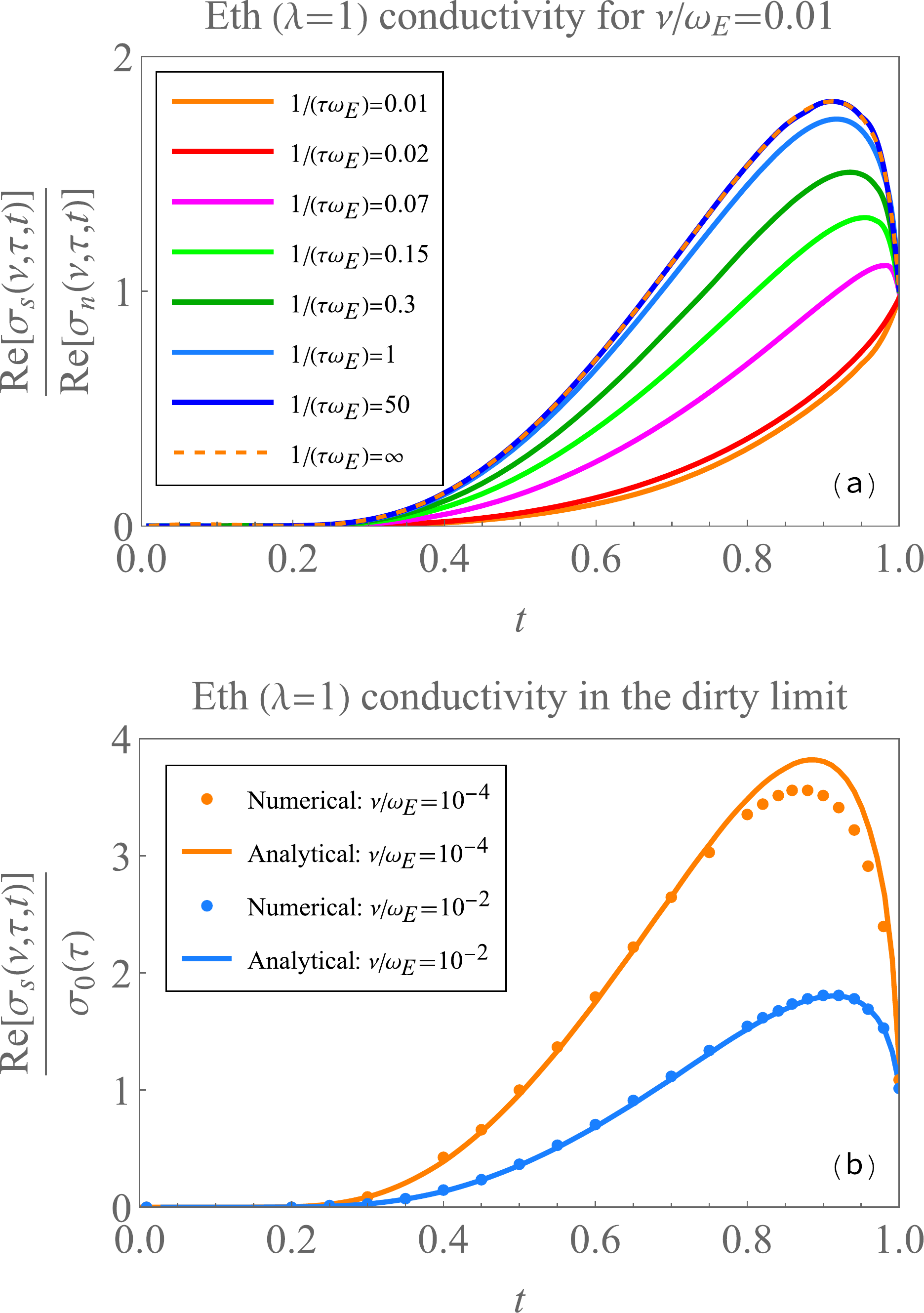}
\caption{Electrical conductivity plots. (a) Normalized ratio of the real part of the superconducting conductivity to the normal-state conductivity ($\sigma_{s}/\sigma_{n}$) in Eliashberg theory (Eth) versus reduced temperature $t\equiv T/T_{c}$ for various inverse scattering values $1/(\tau\omega_{E})$. The numerical results are obtained using Eq.~\eqref{eq:KleinSigma}, with electron-phonon coupling $\lambda=1$, external frequency $\nu/\omega_{E}=0.01$, where $\omega_{E}$ is the Einstein frequency. The bottom curve corresponds to $1/(\tau\omega_{E})=0.01$ (the clean limit) and the top curve is $1/(\tau\omega_{E})=\infty$ (the dirty limit). The value of $1/(\tau\omega_{E})$ increases upwards from the bottom curve to the top curve. (b) The same quantity as in figure (a), but now in the dirty limit $1/\left(\tau\omega_{E}\right)=\infty$. The blue numerical result is the same as in (a), and the analytical blue curve is obtained using Eq.~\eqref{eq:EthDC1}. The orange curves show the numerical and analytical results for a lower frequency $\nu/\omega_{E}=10^{-4}$.}
\label{fig:Figure1}
\end{figure}

\begin{figure}[t]
\centering
\includegraphics[width=7.5cm,height=11cm]{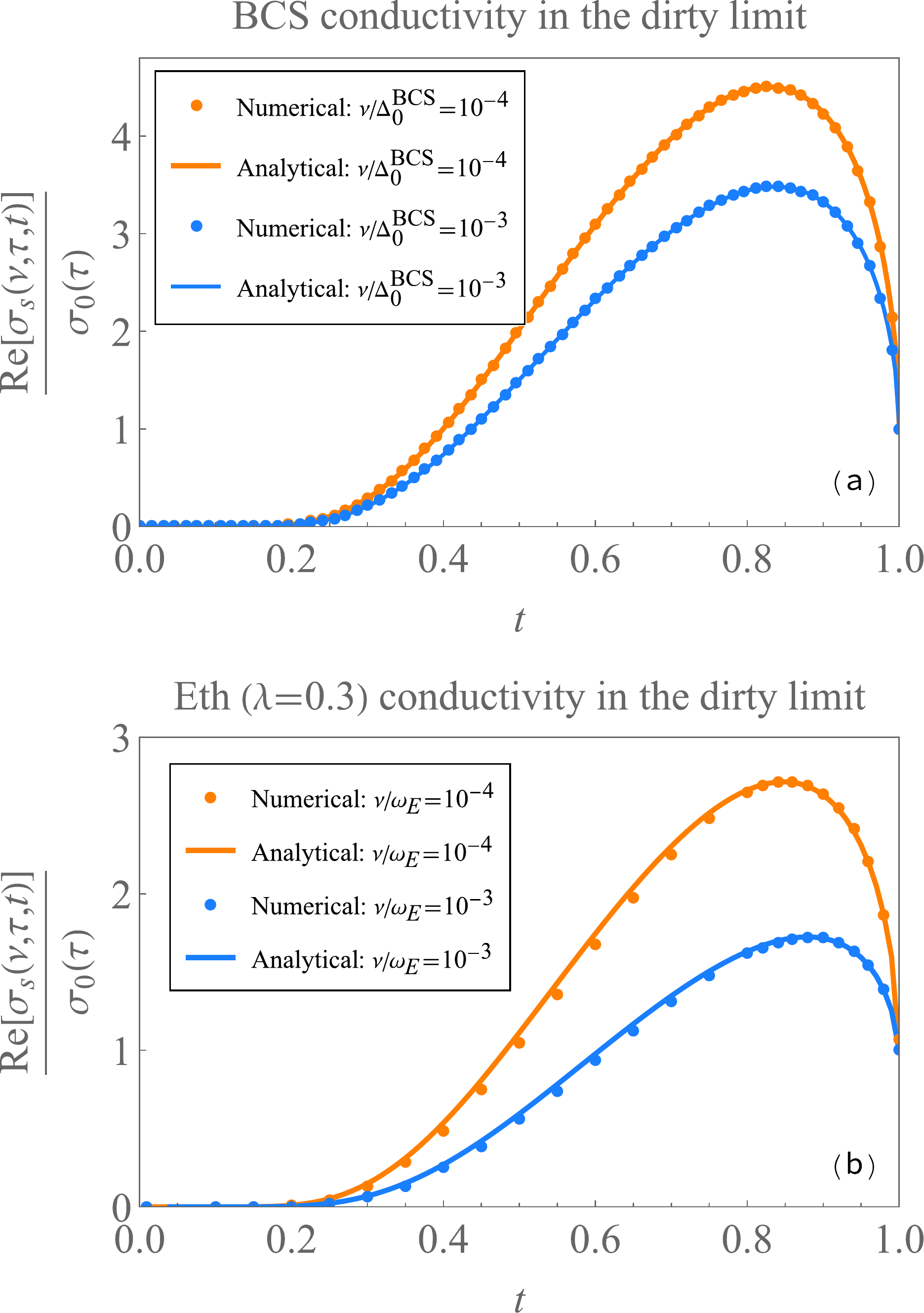}
\caption{The conductivity ratio versus $t\equiv T/T_{c}$ in the dirty limit. (a) BCS theory results. The numerical results are computed using Eq.~\eqref{eq:MattisBardeen} and the solid curves are obtained using Eq.~\eqref{eq:BCSDC}. (b) Eliashberg theory results for $\lambda=0.3$, $T_{c}/\omega_{E}=0.009923$. The numerical results are computed using Eq.~\eqref{eq:DirtySigma} and the solid curves are obtained using Eq.~\eqref{eq:EthDC1}.}
\label{fig:Figure2}
\end{figure}

\begin{figure}[t]
\centering\includegraphics[width=7.5cm,height=11cm]{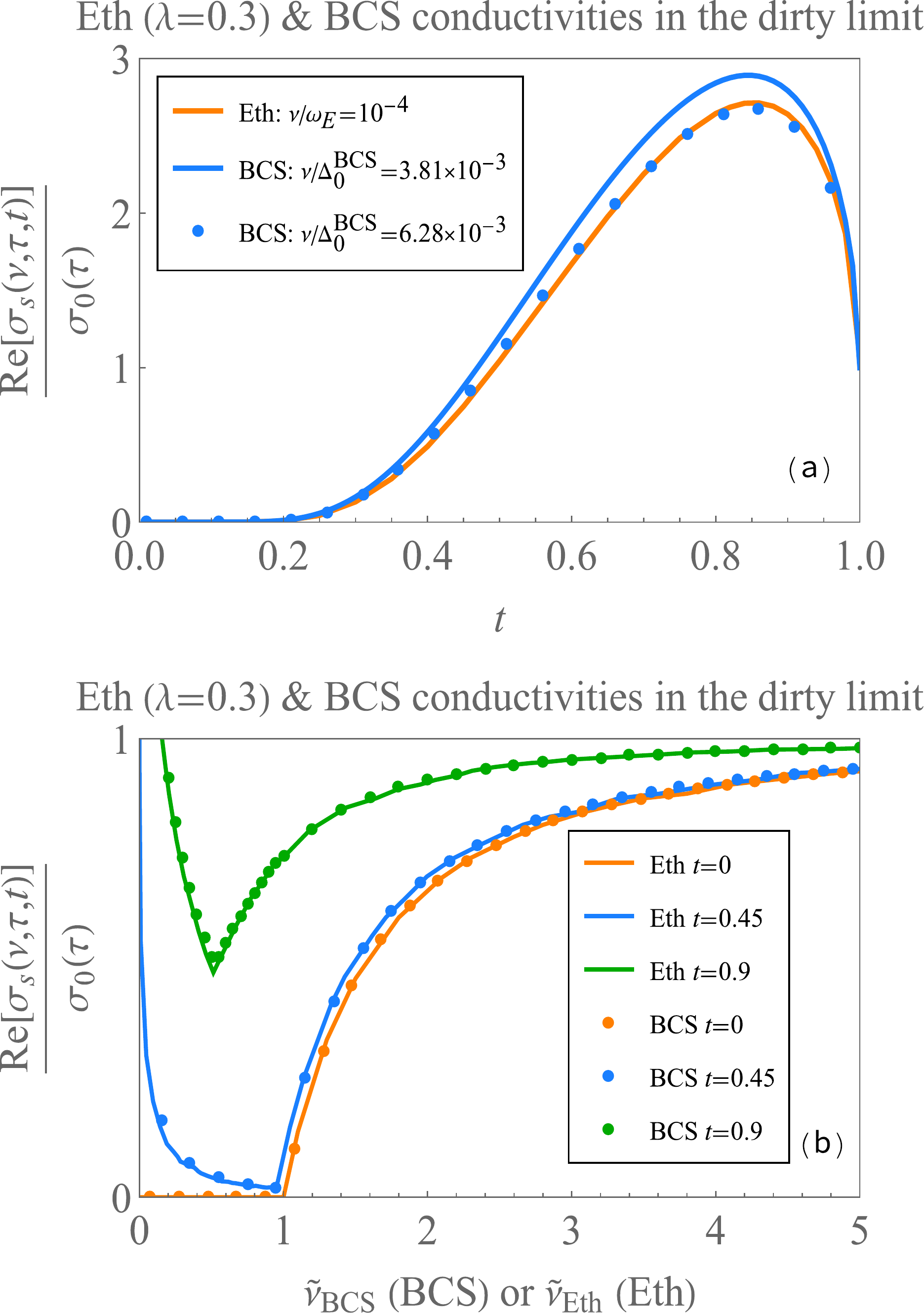}
\caption{Electrical conductivities for the Eliashberg and BCS theories. (a) The Eth ($\lambda=0.3$) and BCS conductivity ratios versus $t\equiv T/T_{c}$ in the dirty limit. The orange curve is for Eth with $\nu/\omega_{E}=10^{-4}$. The blue curve is BCS theory, where we also use $\nu/\omega_{E}=10^{-4}$, which implies $\nu/\Delta^{\BCS}_{0}=(\nu/\omega_{E})(\omega_{E}/\Delta^{\BCS}_{0})\approx3.81\times10^{-3}$. Here we use Eq.~\eqref{eq:GapEdges1} for the BCS gap edge $\Delta^{\BCS}_{0}$. In the case of the blue data points, for the gap we use the weak-coupling result given in Eq.~\eqref{eq:GapEdges2}. This plot shows that, using a BCS gap with a weak-coupling correction of $1/\sqrt{e}$ gives an electrical conductivity that agrees with the small-$\lambda$ Eth result.
(b) The frequency-dependent conductivity in Eth ($\lambda=0.3$) and BCS theory in the dirty limit. The solid curves are for Eth and the points are for BCS theory. Here we define a dimensionless  frequency by $\widetilde{\nu}=\nu/\left(2\Delta_{0}\right)$. Note that we use different horizontal axes for the Eth and BCS curves.}
\label{fig:Figure3}
\end{figure}

In this section, we present our numerical results for the electrical conductivity. For the details on how we compute $\Delta(\omega)$ and $Z(\omega)$, we refer the reader to Ref.~\cite{Mirabi2020} and the references therein. In Fig.~\ref{fig:Figure1}(a), we show the numerical results for the real part of the conductivity, as determined from Eq.~\eqref{eq:KleinSigma}. Here we plot $\textrm{Re}\sigma_{s}/\textrm{Re}\sigma_{n}$, where $\sigma_{n}$ is the normal-state conductivity determined by setting $\Delta=0$ in Eq.~\eqref{eq:KleinSigma}. The parameters we used are $\lambda=1$, $\omega_{E}=1$meV, and $\nu/\omega_{E}=0.01$. The impurity scattering rate ranges from $1/(\tau\omega_{E})=0.01$ (clean limit) to $1/(\tau\omega_{E})=\infty$ (dirty limit). At $1/(\tau\omega_{E})=50$ the dirty limit is already reached. In Fig.~\ref{fig:Figure1}(b), we plot the conductivity in the dirty limit and compare it with the analytical prediction~\eqref{eq:EthDC1}. For $\nu/\omega_{E}=10^{-2}$ we find excellent agreement between our numerical and theoretical results. For $\nu/\omega_{E}=10^{-4}$ there is a slight discrepancy at reduced temperatures $T/T_{c}\sim0.9$. This can be understood from the fact that, in this temperature regime, $|\Delta^{\Eth}_{2}/\omega_{E}|\sim10^{-5}-10^{-4}$, and thus the requirement $|\Delta^{\Eth}_{2}|\ll\nu$ for Eq.~\eqref{eq:EthDC1} to be valid is not met. 

The peak in the conductivity is known as the `Hebel-Slichter' or `coherence' peak, and it arises due to the coherence factors in the response~\cite{Bennemann}. The response function involves a convolution of two single-particle densities of states. The single-particle density of states diverges at the gap edge, and this becomes non-integrable as the external frequency $\nu$ decreases to zero. However, if the frequency is non-zero there is no longer a divergence, and if $\nu$ increases even further, the peak is no longer present. 

In Fig.~\ref{fig:Figure2}(a) we plot the conductivity for BCS theory, in the dirty limit, for two frequencies. For the orange curves $\nu/\Delta^{\BCS}_{0}=10^{-4}$ and for the blue curves $\nu/\Delta^{\BCS}_{0}=10^{-3}$. The numerical results are obtained using the Mattis-Bardeen formula~\eqref{eq:MattisBardeen}, with $\Delta$ given by the temperature-dependent gap function, whereas the solid curves are obtained using our low-frequency expression~\eqref{eq:BCSDC}. In Fig.~\ref{fig:Figure2}(b) we plot the electrical conductivity for Eth, in the dirty limit, for low frequencies. Here we study the weak-coupling limit and set $\lambda=0.3$. For the orange curves $\nu/\omega_{E}=10^{-4}$ and for the blue curves $\nu/\omega_{E}=10^{-3}$. The numerical results are obtained using Eq.~\eqref{eq:DirtySigma}, whereas the solid curves are obtained using our low-frequency expression~\eqref{eq:EthDC1}. For both BCS theory and Eth we observe good agreement between the numerical and analytical results. 

In Fig.~\ref{fig:Figure3}(a) we plot the conductivity ratio for BCS theory and Eth, both in the dirty limit. To make a meaningful comparison between the two theories, here we fix $\nu/\omega_{E}=10^{-4}$. For the BCS curve (solid blue), we determine $\nu/\Delta^{\BCS}_{0}$ using the BCS result in Eq.~\eqref{eq:GapEdges1}: $\nu/\Delta^{\BCS}_{0}=(\nu/\omega_{E})(\omega_{E}/\Delta^{\BCS}_{0})$. For the other BCS result (blue points), we use the weak-coupling result for $\Delta^{\Eth}_{1}$ in Eq.~\eqref{eq:GapEdges2}. The disagreement between the orange and blue curves shows that, for a fixed, comparable frequency, the weak-coupling Eth conductivity is not the same as the BCS conductivity. The agreement between the orange curve and the blue dots shows that, when weak-coupling corrections are incorporated into BCS theory, the result agrees with the conductivity for Eth. This necessarily means that the respective conductivities are normalized using different energy scales, a point alluded to in the previous section. Electrical conductivity may thus be a useful way to determine the correct~\cite{Mirabi2020, Yuzbashyan2022} weak-coupling limit of Eth. In particular, while $\sigma_{0}$ does become smaller in the dirty limit, the normalized conductivity in Fig.~\ref{fig:Figure3}(a) shows a quantitative difference between Eth and BCS theory. This should be observable in an experiment.

In Fig.~\ref{fig:Figure3}(b), we plot the electrical conductivity (in the dirty limit) as a function of the dimensionless frequency $\widetilde{\nu}=\frac{\nu}{2\Delta_{0}}$, for fixed temperatures $t=0, 0.45$, and $t=0.9$. Note that, for Eth we use $\Delta^{\Eth}_{0}$ whereas for BCS theory we use $\Delta^{\BCS}_{0}$ for the normalization. Thus, this plot shows that the Eth and BCS  conductivities have the same functional dependence on the dimensionless frequency; however, because the frequencies have a different scale, the conductivities themselves will differ. This is another way to visualize Eq.~\eqref{eq:EthBCSSigmaComp}.

\subsubsection{NMR relaxation rate}
\label{sec:NMRRelaxRate}

\begin{figure}[h]
\centering
\includegraphics[width=7.5cm,height=5.5cm]{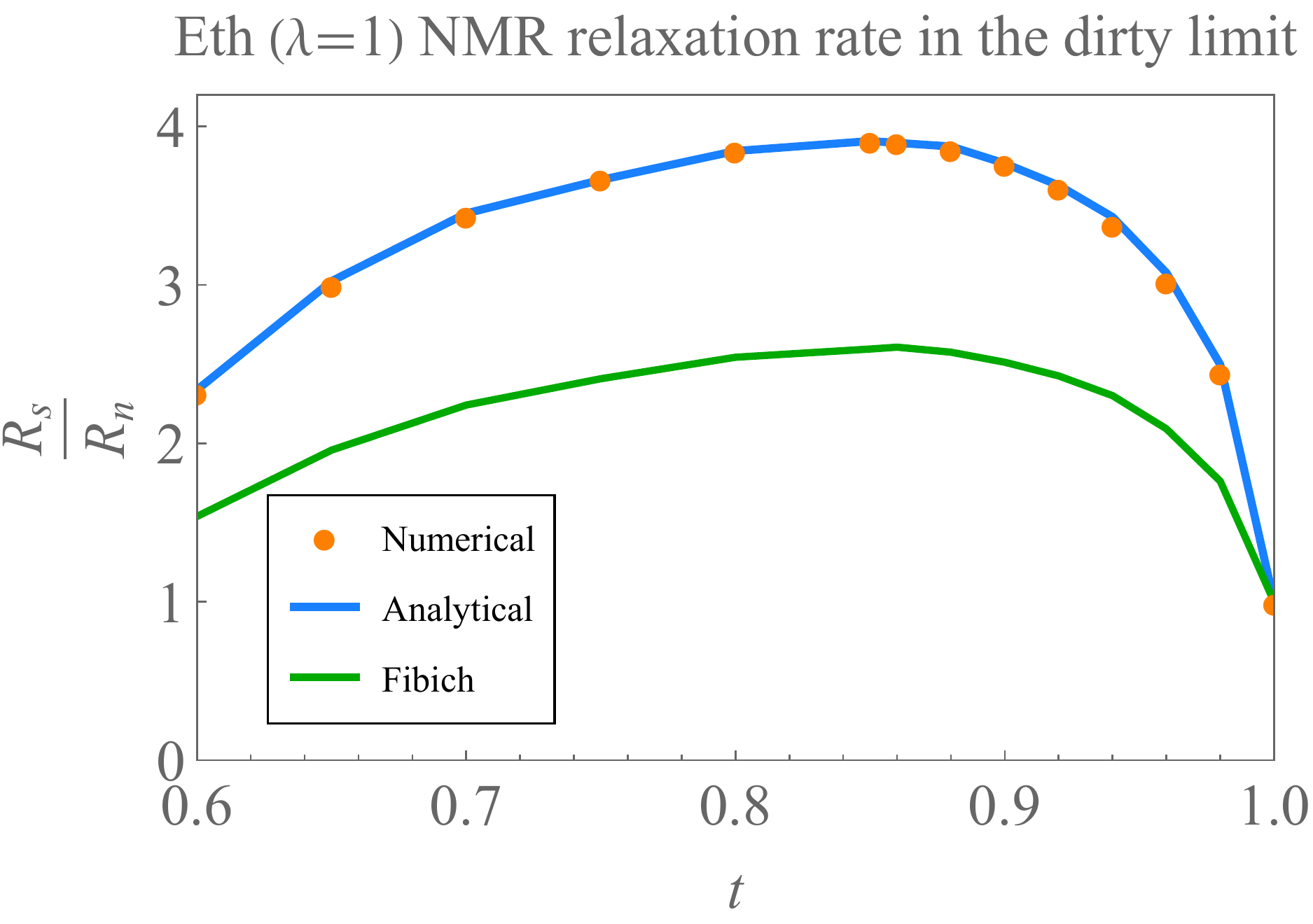}
\caption{The nuclear magnetic resonance (NMR) relaxation rate $R_{s}/R_{n}$ in Eliashberg theory (Eth) versus reduced temperature $t\equiv T/T_{c}$. Here, the electron-phonon coupling strength is $\lambda=1$. The orange points are the numerical results obtained using Eq.~\eqref{eq:EthNMR}, the blue curve is our analytical result~\eqref{eq:EthDC2}, and the green curve is Fibich's result~\eqref{eq:Fibich}. The lines merely connect the data points.}
\label{fig:Figure4}
\end{figure}

We now investigate the NMR relaxation rate for Eth. Fibich~\citep{Fibich1965, Fibich1965b} previously obtained the following result:
\begin{equation}
\label{eq:Fibich}
\left(\frac{R_{s}}{R_{n}}\right)^{\textrm{Fibich}} =2f\left(\Delta_{1}\right)-\Delta_{1}f^{\prime} \left(\Delta_{1}\right)\log\left(\frac{4\Delta_{1}}{\left|\Delta_{2}\right|}\right).
\end{equation}
It is important to note that the detailed analysis that leads to Eq.~\eqref{eq:EthDC2} is quite different than that which would (naively) lead to Eq.~\eqref{eq:Fibich}.
In Fig.~\ref{fig:Figure4}, we compare our analytical result~\eqref{eq:EthDC2}, Fibich's result~\eqref{eq:Fibich}, and the numerical result obtained using the full expression~\eqref{eq:EthNMR} for the case $\lambda=1$. We determined $\Delta_{2}$ by solving the Eth gap equations with a mesh of $10^{-5}\omega_{E}$ and by finding when Eqs.~\eqref{eq:Delta1Def}-\eqref{eq:Delta2Def} were approximately satisfied. Since $\Delta_{2}$ changes rapidly in the vicinity of $\omega=\Delta_{1}$, we used an interpolation method (using three or so values at consecutive frequencies) to deduce the values of $\Delta_{1}$ and $\Delta_{2}$. Of most interest are the heights of the peaks. As pointed out in Ref.~\cite{Parks1}, the coherence peak in the NMR relaxation rate arises from the dynamics of the electron-phonon interaction. Indeed, the height of the coherence peak provides a measure of the magnitude of inelastic scattering in a system~\cite{Marsiglio1994}. As Fig.~\ref{fig:Figure4} shows, our analytical result agrees with the numerical result, whereas Fibich's result is clearly erroneous.

\section{Conclusion}
\label{sec:Conclusion}

The purpose of this article has been to extend the weak-coupling analysis of Eliashberg theory (Eth). Specifically, our goal was to investigate the finite-frequency electrical conductivity and observe how the dynamical nature of the electron-phonon interaction plays a role. By focusing primarily on the dirty limit, we have derived closed-form expressions for the low-frequency conductivity in both BCS theory and Eth. For Eth we found that there are two cases to consider, based on whether the external frequency is much higher or lower than a particular imaginary component of the gap. In the dirty limit, when the external frequency is zero the electrical conductivity is proportional to the NMR relaxation rate. We have performed numerical calculations which agree quite well with our theoretical predictions. Our analytical results corrected an earlier expression presented in the literature, and thus our modified analysis should be important for NMR relaxation rate studies in a bevy of superconductivity applications. 

It has been shown previously that there are dimensionless ratios (for example the normalized specific heat jump) which have ``universal'' values, meaning that in the weak-coupling limit these values tend to their counterpart BCS predictions. For optical conductivity, however, we have argued that there are no such universal values, and consequently our analysis highlights that BCS theory and Eth have observable differences in their optical response in the weak-coupling limit. The importance of this result is that it can provide a method to probe the strength of the electron-phonon interaction in a system.

In regards to future work, determining a closed-form expression for the imaginary gap component $\Delta_{2}$ would be extremely beneficial, particularly since it would enable the NMR relaxation rate to be determined completely analytically as a function of temperature. Another interesting topic includes undertaking analytical studies of the electrical conductivity for Eth in the clean limit. In this case the role of elastic impurities will be important.

\textbf{Acknowledgments}\\
R.B. was supported by the Department of Physics and Astronomy, Dartmouth College, and also by D\'epartement de physique, Universit\'e de Montr\'eal where part of this work was performed. F.M. was supported in part by the Natural Sciences and Engineering Research Council of Canada (NSERC) and by an MIF from the Province of Alberta.

\section{Methods}
\subsection{Low-frequency limit of the Mattis-Bardeen formula}
\label{app:MattisBardeenAnalysis}

In this section we derive Eq.~\eqref{eq:BCSDC} starting from Eq.~\eqref{eq:MattisBardeen}. Here we shall write $\Delta$ for the gap, with the understanding that it denotes the BCS gap. In the limit $0<\nu\ll\Delta$, only the first term in Eq.~\eqref{eq:MattisBardeen} contributes. Using the identity $\tanh\left(\frac{x}{2T}\right)=1-2f(x)$, the tanh functions can be expanded to linear order to obtain 
\begin{align}
\widetilde{\sigma}^{\BCS}_{1} & = 2\int_{\Delta}^{\infty}\frac{\omega\left(\omega+\nu\right)+\Delta^{2}}{\sqrt{\omega^{2}-\Delta^{2}}\sqrt{\left(\omega+\nu\right)^{2}-\Delta^{2}}}\left(-\frac{\partial f}{\partial \omega}\right)d\omega \nonumber \\
& = \int_{\Delta}^{\infty}\frac{4\Delta^{2}}{\sqrt{\omega^{2}-\Delta^{2}}\sqrt{\left(\omega+\nu\right)^{2}-\Delta^{2}}}\left(-\frac{\partial f}{\partial \omega}\right)d\omega  \nonumber\\
&\quad+ 2\int_{\Delta}^{\infty}\left(-\frac{\partial f}{\partial \omega}\right)d\omega + \mathcal{O}\left(\nu\right).
\end{align}
After evaluating the second integral, we have 
\begin{align}
\widetilde{\sigma}^{\BCS}_{1} & = 2f\left(\Delta\right)-4\Delta^{2}\int_{\Delta}^{\infty}\frac{f^{\prime}\left(\omega\right)}{\sqrt{\omega^{2}-\Delta^{2}}\sqrt{\left(\omega+\nu\right)^{2}-\Delta^{2}}}d\omega \nonumber \\
 & =  -4\Delta^{2}f^{\prime}\left(\Delta\right)\int_{\Delta}^{\infty}\frac{1}{\sqrt{\omega^{2}-\Delta^{2}}\sqrt{\left(\omega+\nu\right)^{2}-\Delta^{2}}}d\omega \nonumber \\
  & \quad -4\Delta^{2}\int_{\Delta}^{\infty}\frac{f^{\prime}\left(\omega\right)-f^{\prime}\left(\Delta\right)}{\sqrt{\omega^{2}-\Delta^{2}}\sqrt{\left(\omega+\nu\right)^{2}-\Delta^{2}}}d\omega \nonumber\\
 &\quad + 2f\left(\Delta\right).
\end{align}
Since the second integral is well behaved as $\nu/\Delta\rightarrow0$, we can set $\nu=0$ in this integral. In the first integral, substitute $\omega=x-\frac{\nu}{2}$ to obtain
\begin{align}
I_{1} & = \int_{\Delta}^{\infty}\frac{1}{\sqrt{\omega^{2}-\Delta^{2}}\sqrt{\left(\omega+\nu\right)^{2}-\Delta^{2}}} d\omega \nonumber \\
 & = \int_{\Delta+\frac{\nu}{2}}^{\infty}\frac{1}{\sqrt{\left(x-\frac{\nu}{2}\right)^{2}-\Delta^{2}}\sqrt{\left(x+\frac{\nu}{2}\right)^{2}-\Delta^{2}}} dx,\ x=\frac{1}{y}\nonumber \\
 & = \int_{0}^{\frac{1}{\Delta+\frac{\nu}{2}}}\frac{1}{\sqrt{\left(1-y\frac{\nu}{2}\right)^{2}-y^{2}\Delta^{2}}\sqrt{\left(1+y\frac{\nu}{2}\right)^{2}-y^{2}\Delta^{2}}} dy.
\end{align}
Now substitute $y=\frac{1}{\Delta+\frac{\nu}{2}}t$:
\begin{align}
I_{1} & = \frac{1}{\Delta+\frac{\nu}{2}}\int_{0}^{1}\biggl\{\frac{1}{\sqrt{\left(1-\frac{t}{\Delta+\frac{\nu}{2}}\frac{\nu}{2}\right)^{2}-\left(\frac{\Delta t}{\Delta+\frac{\nu}{2}}\right)^{2}}} \nonumber\\
&\hspace{1cm} \times\frac{1}{\sqrt{\left(1+\frac{t}{\Delta+\frac{\nu}{2}}\frac{\nu}{2}\right)^{2}-\left(\frac{\Delta t}{\Delta+\frac{\nu}{2}}\right)^{2}}}\biggr\} dt \nonumber \\
 & = \frac{2}{2\Delta+\nu}\int_{0}^{1}\frac{1}{\sqrt{1-t^{2}}\sqrt{1-\left(\frac{2\Delta-\nu}{2\Delta+\nu}\right)^{2}t^{2}}} dt.
\end{align}
The complete elliptic integral of the first kind is defined by (see
pg.~501 of Ref.~\citep{Whittaker_WatsonBook} or Eq.~19.2.8 of Ref.~\citep{NIST2020}):
\begin{equation}
K\left(k\right)=\int_{0}^{1}\frac{1}{\sqrt{1-t^{2}}\sqrt{1-k^{2}t^{2}}} dt.
\end{equation}
Thus, we find that the integral is given by 
\begin{align}
I_{1}&=\int_{\Delta}^{\infty}\frac{1}{\sqrt{\omega^{2}-\Delta^{2}}\sqrt{\left(\omega+\nu\right)^{2}-\Delta^{2}}} d\omega \nonumber\\
&=\frac{2}{2\Delta+\nu}K\left(\frac{2\Delta-\nu}{2\Delta+\nu}\right).
\end{align}
As a result, the conductivity now becomes 
\begin{align}
\widetilde{\sigma}^{\BCS}_{1} &= 2f\left(\Delta\right) -4\Delta^{2}f^{\prime} \left(\Delta\right)\frac{2}{2\Delta+\nu}K\left(\frac{2\Delta-\nu}{2\Delta+\nu}\right) \nonumber\\
&\quad -4\Delta^{2}\int_{\Delta}^{\infty}\frac{f^{\prime}\left(\omega\right)-f^{\prime}\left(\Delta\right)}{\omega^{2}-\Delta^{2}} d\omega .
\end{align}
To take the small-$\nu$ limit, we use the following result (see pg. 521 of Ref.~\citep{Whittaker_WatsonBook}):
\begin{equation}
\lim_{k\rightarrow0}K\left(\sqrt{1-k^{2}}\right)=\log\left(\frac{4}{k}\right).
\end{equation}
Using this result, we find 
\begin{equation}
\lim_{\frac{\nu}{\Delta_{0}}\rightarrow0}\frac{2}{2\Delta+\nu}K\left(\frac{2\Delta-\nu}{2\Delta+\nu}\right)=\frac{1}{2\Delta}\log\left(\frac{8\Delta}{\nu}\right).
\end{equation}
Thus, the real part of the conductivity is given by
\begin{align}
\widetilde{\sigma}^{\BCS}_{1} &= 2f\left(\Delta\right)-2\Delta f^{\prime}\left(\Delta\right)\log\left(\frac{8\Delta}{\nu}\right) \nonumber\\
&\quad +4\Delta\int_{1}^{\infty}\frac{f^{\prime}\left(\Delta\right)-f^{\prime}\left(\Delta x\right)}{x^{2}-1} dx.
\end{align}

\subsection{Low-frequency limit for dirty Eliashberg superconductors}
\label{app:EthAnalysis}

In this section we derive Eq.~\eqref{eq:EthDC2}. Our starting point for the low-frequency analysis is Eq.~\eqref{eq:EthNMR}:
\begin{align}
\left(\frac{R_{s}}{R_{n}}\right)^{\Eth} &= 2\int_{0}^{\infty}\biggl\{\left[\mathrm{Re}\frac{\omega}{\sqrt{\omega^{2}-\Delta^{2}}}\right]^{2} \nonumber\\ 
&\quad\quad +\left[\mathrm{Re}\frac{\Delta}{\sqrt{\omega^{2}-\Delta^{2}}}\right]^{2}\biggr\} \left(-\frac{\partial f}{\partial\omega}\right) d\omega.
\end{align}
Following Fibich~\citep{Fibich1965,Fibich1965b}, we let $\Delta_{1}$ and $\Delta_{2}$ be defined as in Eqs.~\eqref{eq:Delta1Def}-\eqref{eq:Delta2Def}, and we shall assume that $|\Delta_{2}|\ll\Delta_{1}$. The main contribution to the integral comes from the branch cut. Let $\Delta\approx\Delta_{1}+i\Delta_{2}$ so that
\begin{align}
\omega^{2}-\Delta{}^{2} & = \omega^{2}-\left(\Delta_{1}^{2}-\Delta_{2}^{2}+2i\Delta_{1}\Delta_{2}\right)\nonumber \\
 & = \sqrt{\left|\left(\omega^{2}-\Delta_{1}^{2}+\Delta_{2}^{2}\right)^{2}+4\Delta_{1}^{2}\Delta_{2}^{2}\right|}e^{i\theta}.
\end{align}
The square root of $\omega^{2}-\Delta^{2}$ is defined to have a real part that has the same sign as $\omega$. Thus, in the limit $\Delta_{2}\rightarrow0$ this requires $\omega\geq\Delta_{1}$. Hence, we integrate from $\omega=\Delta_{1}$ to $\omega=\infty$.  The angle $\theta$ is then given by 
\begin{equation}
\theta=\arctan\frac{2\Delta_{1}\left|\Delta_{2}\right|}{\left|\omega^{2}-\Delta_{1}^{2}+\Delta_{2}^{2}\right|}.
\end{equation}
Here we used the fact that $\Delta_{2}<0$. After performing some algebraic and trigonometric manipulations, we find 
\begin{align}
 &  \left[\mathrm{Re}\frac{\omega}{\sqrt{\omega^{2}-\Delta^{2}}}\right]^{2}+\left[\mathrm{Re}\frac{\Delta\left(\omega\right)}{\sqrt{\omega^{2}-\Delta^{2}}}\right]^{2}\nonumber \\
 & = 1+\frac{\omega^{2}+\Delta_{1}^{2}+\Delta_{2}^{2}-\sqrt{\left(\omega^{2}-\Delta_{1}^{2}+\Delta_{2}^{2}\right)^{2}+4\Delta_{1}^{2}\Delta_{2}^{2}}}
 {2\sqrt{\left(\omega^{2}-\Delta_{1}^{2}+\Delta_{2}^{2}\right)^{2}+4\Delta_{1}^{2}\Delta_{2}^{2}}} \nonumber\\
& \quad+\frac{\left(\Delta_{1}^{2}-\Delta_{2}^{2}\right)\omega^{2}-\left(\Delta_{1}^{2}+\Delta_{2}^{2}\right)^{2}}{\left(\omega^{2}-\Delta_{1}^{2}+\Delta_{2}^{2}\right)^{2}+4\Delta_{1}^{2}\Delta_{2}^{2}}.
 \label{eq:FibichInt}
\end{align}
Note that this corrects the typo in Eq.~(8) of Ref.~\citep{Fibich1965}. 
For the first term in Eq.~\eqref{eq:FibichInt}, we obtain
\begin{equation}
I_{1} = 2\int_{\Delta_{1}}^{\infty}\left(-\frac{\partial f}{\partial\omega}\right) d\omega = 2f\left(\Delta_{1}\right).
\end{equation}

Now let us focus on the second term in Eq.~\eqref{eq:FibichInt}. Let $U=\omega^{2}+\Delta_{1}^{2}+\Delta_{2}^{2}, V=\left(\omega^{2}-\Delta_{1}^{2}+\Delta_{2}^{2}\right)$. We then have
\begin{align}
I_{2} & = 2\int\frac{U-\sqrt{V^{2}+4\Delta_{1}^{2}\Delta_{2}^{2}}}{2\sqrt{V^{2}+4\Delta_{1}^{2}\Delta_{2}^{2}}}\left(-\frac{\partial f}{\partial\omega}\right) d\omega \nonumber \\
 & = -f^{\prime}\left(\Delta_{1}\right)\int \frac{U - \sqrt{V^{2}+4\Delta_{1}^{2}\Delta_{2}^{2}}}{\sqrt{V^{2}+4\Delta_{1}^{2}\Delta_{2}^{2}}} d\omega \nonumber \\
 &\quad +\int\left[f^{\prime}\left(\Delta_{1}\right)-f^{\prime}\left(\omega\right)\right]\frac{U - \sqrt{V^{2}+4\Delta_{1}^{2}\Delta_{2}^{2}}}{\sqrt{V^{2}+4\Delta_{1}^{2}\Delta_{2}^{2}}} d\omega.\label{eq:FibichInt2}
\end{align}
The first integral can be analyzed as follows; substitute $y=\omega^{2}-\Delta_{1}^{2}+\Delta_{2}^{2}$ to obtain
\begin{align}
I_{2i} & \approx -f^{\prime}\left(\Delta_{1}\right)\int_{\Delta_{1}}^{\infty}\frac{2\Delta_{1}^{2}}{\sqrt{\left(\omega^{2}-\Delta_{1}^{2}+\Delta_{2}^{2}\right)^{2}+4\Delta_{1}^{2}\Delta_{2}^{2}}} d\omega \nonumber \\
 & = -\Delta_{1}^{2}f^{\prime}\left(\Delta_{1}\right)\int_{\Delta_{2}^{2}}^{\infty}\frac{1}{\sqrt{y+\Delta_{1}^{2}-\Delta_{2}^{2}}}\frac{1}{\sqrt{y^{2}+4\Delta_{1}^{2}\Delta_{2}^{2}}} dy \nonumber \\
 & \approx -\Delta_{1}^{2}f^{\prime}\left(\Delta_{1}\right)\int_{0}^{\infty}\frac{1}{\sqrt{y+\Delta_{1}^{2}}}\frac{1}{\sqrt{y^{2}+4\Delta_{1}^{2}\Delta_{2}^{2}}} dy.
\end{align}
Now let $y=x\Delta_{1}^{2}$. We then have
\begin{align}
I_{2i} &= -\Delta_{1}f^{\prime}\left(\Delta_{1}\right)\int_{0}^{\infty}\frac{1}{\sqrt{x+1}}\frac{1}{\sqrt{x^{2}+\left(\frac{2\Delta_{2}}{\Delta_{1}}\right)^{2}}} dx \nonumber \\
 &= \left.\frac{-\Delta_{1}f^{\prime}\left(\Delta_{1}\right)}{\sqrt{x+1}}\log\left[x+\sqrt{x^{2}+\left(\frac{2\Delta_{2}}{\Delta_{1}}\right)^{2}}\right]\right|_{x=0}^{x=\infty}\nonumber \\
 & \quad -\frac{1}{2}\Delta_{1}f^{\prime}\left(\Delta_{1}\right)\int_{0}^{\infty}\frac{\log\left[x+\sqrt{x^{2}+\left(\frac{2\Delta_{2}}{\Delta_{1}}\right)^{2}}\right]}{\left(x+1\right)^{\frac{3}{2}}} dx.
\end{align}
Now make further simplifications to obtain 
\begin{align}
I_{2i} &\rightarrow \Delta_{1}f^{\prime}\left(\Delta_{1}\right)\log\left|\frac{2\Delta_{2}}{\Delta_{1}}\right| \nonumber\\
&\quad-\frac{1}{2}\Delta_{1} f^{\prime}\left(\Delta_{1}\right)\int_{0}^{\infty}\frac{\log\left(2x\right)}{\left(x+1\right)^{\frac{3}{2}}} dx \nonumber\\
&= -\Delta_{1}f^{\prime} \left(\Delta_{1}\right)\log\left(\frac{4\Delta_{1}}{\left|\Delta_{2}\right|}\right).
\end{align}
For the second integral in Eq.~\eqref{eq:FibichInt2}, we obtain
\begin{eqnarray}
I_{2ii} & = & \int_{\Delta_{1}}^{\infty}\left[f^{\prime}\left(\Delta_{1}\right)-f^{\prime}\left(\omega\right)\right] \frac{U - \sqrt{V^{2}+4\Delta_{1}^{2}\Delta_{2}^{2}}}{\sqrt{V^{2}+4\Delta_{1}^{2}\Delta_{2}^{2}}} d\omega \nonumber \\
 & \rightarrow & 2\Delta_{1}^{2}\int_{\Delta_{1}}^{\infty}\frac{f^{\prime}\left(\Delta_{1}\right)-f^{\prime}\left(\omega\right)}{\omega^{2}-\Delta_{1}^{2}} d\omega.
\end{eqnarray}

Let us focus on the third term in Eq.~\eqref{eq:FibichInt}:
\begin{align}
I_{3} & = 2\int\frac{\left(\Delta_{1}^{2}-\Delta_{2}^{2}\right)\omega^{2}-\left(\Delta_{1}^{2}+\Delta_{2}^{2}\right)^{2}}{\left(\omega^{2}-\Delta_{1}^{2}+\Delta_{2}^{2}\right)^{2}+4\Delta_{1}^{2}\Delta_{2}^{2}}\left(-\frac{\partial f}{\partial\omega}\right) d\omega \nonumber \\
 & = -2f^{\prime}\left(\Delta_{1}\right)\int \frac{\left(\Delta_{1}^{2}-\Delta_{2}^{2}\right)\omega^{2}-\left(\Delta_{1}^{2}+\Delta_{2}^{2}\right)^{2}}{\left(\omega^{2}-\Delta_{1}^{2}+\Delta_{2}^{2}\right)^{2}+4\Delta_{1}^{2}\Delta_{2}^{2}} d\omega \nonumber \\
 &\quad + 2\int_{\Delta_{1}}^{\infty}\biggl\{\frac{\left(\Delta_{1}^{2}-\Delta_{2}^{2}\right)\omega^{2}-\left(\Delta_{1}^{2}+\Delta_{2}^{2}\right)^{2}}{\left(\omega^{2}-\Delta_{1}^{2}+\Delta_{2}^{2}\right)^{2}+4\Delta_{1}^{2}\Delta_{2}^{2}} \nonumber\\
&\hspace{1.5cm} \times \left[f^{\prime}\left(\Delta_{1}\right)-f^{\prime}\left(\omega\right)\right] \biggr\} d\omega.
\label{eq:FibichInt3}
\end{align}
The first integral in Eq.~\eqref{eq:FibichInt3} can be performed exactly:
\begin{align}
I_{3i} &= -2f^{\prime}\left(\Delta_{1}\right)\int \frac{\left(\Delta_{1}^{2}-\Delta_{2}^{2}\right)\omega^{2}-\left(\Delta_{1}^{2}+\Delta_{2}^{2}\right)^{2}}{\left(\omega^{2}-\Delta_{1}^{2}+\Delta_{2}^{2}\right)^{2}+4\Delta_{1}^{2}\Delta_{2}^{2}} d\omega\nonumber \\
 &=  -2f^{\prime}\left(\Delta_{1}\right)\biggl\{ \frac{1}{4}\Delta_{1}\log\left[\frac{\left(\Delta_{1}-\omega\right)^{2}+\Delta_{2}^{2}}{\left(\Delta_{1}+\omega\right)^{2}+\Delta_{2}^{2}}\right] \nonumber\\
 &\quad -\frac{1}{2}\Delta_{2}\left[\arctan\left(\frac{\omega+\Delta_{1}}{\Delta_{2}}\right)+\arctan\left(\frac{\omega-\Delta_{1}}{\Delta_{2}}\right)\right]\biggr\} .
\end{align}
Evaluating the definite integral gives:
\begin{align}
I_{3i} &=  -2f^{\prime}\left(\Delta_{1}\right)\int_{\Delta_{1}}^{\infty}\frac{\left(\Delta_{1}^{2}-\Delta_{2}^{2}\right)\omega^{2}-\left(\Delta_{1}^{2}+\Delta_{2}^{2}\right)^{2}}{\left(\omega^{2}-\Delta_{1}^{2}+\Delta_{2}^{2}\right)^{2}+4\Delta_{1}^{2}\Delta_{2}^{2}} d\omega\nonumber \\
 &= -2f^{\prime}\left(\Delta_{1}\right)\biggl\{ \frac{1}{4}\Delta_{1}\log\left(\frac{4\Delta_{1}^{2}+\Delta_{2}^{2}}{\Delta_{2}^{2}}\right) \nonumber\\
&\quad+\frac{1}{2}\Delta_{2}\arctan\left(2\frac{\Delta_{1}}{\Delta_{2}}\right)-\frac{\pi}{2}\left|\Delta_{2}\right|\biggr\} .
\end{align}
For small $\Delta_{2}/\Delta_{1}$, this expression limits to 
\begin{equation}
I_{3i}\rightarrow-\Delta_{1}f^{\prime}\left(\Delta_{1}\right)\log\left(\frac{2\Delta_{1}}{\left|\Delta_{2}\right|}\right).
\end{equation}
For the second integral in Eq.~\eqref{eq:FibichInt3} we have 
\begin{align}
I_{3ii} &= 2\int_{\Delta_{1}}^{\infty}\biggl\{\frac{\left(\Delta_{1}^{2}-\Delta_{2}^{2}\right)\omega^{2}-\left(\Delta_{1}^{2}+\Delta_{2}^{2}\right)^{2}}{\left(\omega^{2}-\Delta_{1}^{2}+\Delta_{2}^{2}\right)^{2}+4\Delta_{1}^{2}\Delta_{2}^{2}} \nonumber\\
&\quad \times \left[f^{\prime}\left(\Delta_{1}\right)-f^{\prime}\left(\omega\right)\right] \biggr\} d\omega \nonumber \\
 &\rightarrow 2\Delta_{1}^{2}\int_{\Delta_{1}}^{\infty}\frac{f^{\prime}\left(\Delta_{1}\right)-f^{\prime}\left(\omega\right)}{\omega^{2}-\Delta_{1}^{2}} d\omega.
\end{align}

Finally, let us summarize our analysis: in the small $|\Delta_{2}|/\Delta_{1}$ limit, we obtain 
\begin{align}
\label{eq:EthDirtySigmaDC3}
\left(\frac{R_{s}}{R_{n}}\right)^{\Eth} & =  {\ensuremath{2f\left(\Delta_{1}\right)}} -\Delta_{1}f^{\prime}\left(\Delta_{1}\right)\log\left(\frac{4\Delta_{1}}{\left|\Delta_{2}\right|}\right) \nonumber\\
&\quad -\Delta_{1}f^{\prime}\left(\Delta_{1}\right)\log\left(\frac{2\Delta_{1}}{\left|\Delta_{2}\right|}\right) \nonumber\\
&\quad +4\Delta_{1}^{2}\int_{\Delta_{1}}^{\infty}\frac{f^{\prime}\left(\Delta_{1}\right)-f^{\prime}\left(\omega\right)}{\omega^{2}-\Delta_{1}^{2}} d\omega  \\
 & = 2f\left(\Delta_{1}\right) -2\Delta_{1}f^{\prime}\left(\Delta_{1}\right)\log\left(\frac{\sqrt{8}\Delta_{1}}{\left|\Delta_{2}\right|}\right) \nonumber\\
 & \quad +4\Delta_{1}\int_{1}^{\infty}\frac{f^{\prime}\left(\Delta_{1}\right)-f^{\prime}\left(x\Delta_{1}\right)}{x^{2}-1} dx.
\end{align}
Interestingly, Fibich's result in Eq.~\eqref{eq:Fibich} is given by the first two terms in Eq.~\eqref{eq:EthDirtySigmaDC3}. It is conceivable that his analysis included only the first two terms in Eq.~\eqref{eq:FibichInt}.

\subsection{A summary of results for the BCS electrical conductivity}

\subsubsection{General temperatures}

For convenience, here we present a summary of the main results, already in the literature, for the BCS electrical conductivity. In the case where $\Delta$ is frequency independent, the general expression
in Eq.~\eqref{eq:KleinSigma} can be written in a simpler form~\citep{Zimmerman1991}:
\begin{equation}
\frac{\sigma^{\textrm{BCS}}\left(\nu\right)}{\sigma_{0}}=\frac{i}{2\nu\tau}\left(J+\int_{\Delta}^{\infty}I_{2}d\omega\right),
\label{eq:ZimmEq}
\end{equation}
where $\sigma_{0}\equiv ne^{2}\tau/m$ and 
\begin{align}
J\left(\nu\leq2\Delta\right) & =  \int_{\Delta}^{\Delta+\nu}I_{1}d\omega.\\
J\left(\nu\geq2\Delta\right) & =  \int_{\Delta}^{\nu-\Delta}I_{3}d\omega+\int_{\nu-\Delta}^{\nu+\Delta}I_{1}d\omega.
\end{align}
The expressions for $I_{1}, I_{2},$ and $I_{3}$ are given by
\begin{align}
I_{1} & =  \tanh\left(\frac{\omega}{2T}\right)\left\{ \left[1-\frac{\Delta^{2}+\left(\omega-\nu\right)\omega}{P_{4}P_{2}}\right]\frac{1}{P_{4}+P_{2}+i/\tau}\right.\nonumber \\
 & \quad  -\left.\left[1+\frac{\Delta^{2}+\left(\omega-\nu\right)\omega}{P_{4}P_{2}}\right]\frac{1}{P_{4}-P_{2}+i/\tau}\right\} .\\
I_{2} & =  \tanh\left(\frac{\omega+\nu}{2T}\right)\left\{ \left[1+\frac{\Delta^{2}+\omega\left(\omega+\nu\right)}{P_{1}P_{2}}\right]\frac{1}{P_{1}-P_{2}+i/\tau}\right.\nonumber \\
 &  \quad -\left.\left[1-\frac{\Delta^{2}+\omega\left(\omega+\nu\right)}{P_{1}P_{2}}\right]\frac{1}{-P_{1}-P_{2}+i/\tau}\right\} \nonumber \\
 & \quad +\tanh\left(\frac{\omega}{2T}\right)\left\{ \left[1-\frac{\Delta^{2}+\omega\left(\omega+\nu\right)}{P_{1}P_{2}}\right]\frac{1}{P_{1}+P_{2}+i/\tau}\right.\nonumber \\
 & \quad -\left.\left[1+\frac{\Delta^{2}+\omega\left(\omega+\nu\right)}{P_{1}P_{2}}\right]\frac{1}{P_{1}-P_{2}+i/\tau}\right\} .\\
I_{3} & = \tanh\left(\frac{\omega}{2T}\right)\left\{ \left[1-\frac{\Delta^{2}+\left(\omega-\nu\right)\omega}{P_{3}P_{2}}\right]\frac{1}{P_{3}+P_{2}+i/\tau}\right.\nonumber \\
 & \quad -\left.\left[1+\frac{\Delta^{2}+\left(\omega-\nu\right)\omega}{P_{3}P_{2}}\right]\frac{1}{P_{3}-P_{2}+i/\tau}\right\} .
\end{align}
The functions $P_{1}, P_{2}, P_{3}$, and $P_{4}$ are given by 
\begin{align}
P_{1}=\sqrt{\left(\omega+\nu\right)^{2}-\Delta^{2}} &; \quad P_{2}=\sqrt{\omega^{2}-\Delta^{2}}.\\
P_{3}=\sqrt{\left(\omega-\nu\right)^{2}-\Delta^{2}} &; \quad P_{4}=i\sqrt{\Delta^{2}-\left(\omega-\nu\right)^{2}}.
\end{align}

\subsubsection{Dirty limit}
\label{app:DirtyLimitBCS}

We define $\widetilde{\sigma}_{1}(\nu)\equiv\mathrm{Re}\left[\sigma\left(\nu\right)\right]/\sigma_{0}$ and $\widetilde{\sigma}_{2}(\nu)\equiv\mathrm{Im}\left[\sigma\left(\nu\right)\right]/\sigma_{0}$.
For non-zero temperatures $T>0$, the real part of the BCS electrical conductivity can be expressed as~\citep{Mattis1958}: 
\begin{align}
\label{eq:MB}
\widetilde{\sigma}_{1}^{\BCS} &=  \frac{1}{\nu}\int_{\Delta}^{\infty}\biggl\{\frac{\omega\left(\omega+\nu\right)+\Delta^{2}}{\sqrt{\omega^{2}-\Delta^{2}}\sqrt{\left(\omega+\nu\right)^{2}-\Delta^{2}}} \nonumber\\
&\quad \times \left[\tanh\left(\frac{\omega+\nu}{2T}\right)-\tanh\left(\frac{\omega}{2T}\right)\right]\biggr\}
d\omega\nonumber \\
 & \quad -\frac{1}{\nu}\int_{\Delta-\nu}^{-\Delta}\biggl\{
  \frac{\omega\left(\omega+\nu\right)+\Delta^{2}}{\left|\sqrt{\omega^{2}-\Delta^{2}}\right|\sqrt{\left(\omega+\nu\right)^{2}-\Delta^{2}}} \nonumber\\ 
&\quad \times  \Theta\left(\nu-2\Delta\right)\tanh\left(\frac{\omega+\nu}{2T}\right)\biggr\}
 d\omega.
\end{align}
The imaginary part can also be computed; see Ref.~\citep{Mattis1958}.

At $T=0$, we have $f\left(x\right)=\Theta\left(-x\right)$, where $f\left(x\right)$ denotes the Fermi-Dirac distribution function. Since $\Delta_{0}\geq0$, the first term in Eq.~\eqref{eq:MB} does not contribute at $T=0$. Thus, we get 
\begin{equation}
\widetilde{\sigma}_{1}^{\textrm{BCS}} =-\frac{1}{\nu}\int_{\Delta_{0}-\nu}^{-\Delta_{0}}\frac{\Theta\left(\nu-2\Delta_{0}\right)\left[\omega\left(\omega+\nu\right)+\Delta_{0}^{2}\right]}{\sqrt{\left(\omega+\nu\right)^{2}-\Delta_{0}^{2}}\left|\sqrt{\omega^{2}-\Delta_{0}^{2}}\right|}d\omega.
\end{equation}
This can be evaluated as follows. Define $k=\left|\frac{\nu-2\Delta_{0}}{\nu+2\Delta_{0}}\right|$ and substitute $\omega=\left(\frac{\nu}{2}-\Delta_{0}\right)t-\frac{\nu}{2}$,
$d\omega=\left(\frac{\nu}{2}-\Delta_{0}\right)dt$:
\begin{align}
&\widetilde{\sigma}_{1}^{\textrm{BCS}}\left(\nu;T=0\right)  = -\Theta\left(\nu-2\Delta_{0}\right)\frac{\left(\frac{\nu}{2}-\Delta_{0}\right)}{\nu} \nonumber\\
&\quad \times \int_{-1}^{1}\frac{\left(\Delta_{0}-\frac{\nu}{2}\right)^{2}\left(t^{2}-k^{-1}\right)}{\sqrt{\left(\Delta_{0}-\frac{\nu}{2}\right)^{2}\left(\Delta_{0}+\frac{\nu}{2}\right)^{2}\left(1-t^{2}\right)\left(1-k^{2}t^{2}\right)}}dt\nonumber \\
 & \hspace{2.4cm} = \Theta\left(\nu-2\Delta_{0}\right)\left(1-\frac{2\Delta_{0}}{\nu}\right)\nonumber\\ 
 &\hspace{2.7cm}\times\int_{0}^{1}\frac{1-kt^{2}}{\sqrt{\left(1-t^{2}\right)\left(1-k^{2}t^{2}\right)}}dt.
\end{align}

The complete elliptic integrals of the first and second kind are (see Eqs.~19.2.5 and 19.2.8 of Ref.~\citep{NIST2020}):
\begin{align}
K\left(k\right) &= \int_{0}^{1}\frac{1}{\sqrt{1-t^{2}}\sqrt{1-k^{2}t^{2}}}dt. \\
E\left(k\right) &= \int_{0}^{1}\frac{\sqrt{1-k^{2}t^{2}}}{\sqrt{1-t^{2}}}dt.
\end{align}
In terms of these complete elliptic integrals, the zero-temperature conductivity in BCS theory becomes 
\begin{equation}
\widetilde{\sigma}_{1}^{\textrm{BCS}}\left(\nu;T=0\right)=\frac{2\Theta\left(\nu-2\Delta_{0}\right)}{1+k}\left[E\left(k\right)-\left(1-k\right)K\left(k\right)\right].
\end{equation}
Using the definition of $k$ in terms of $\nu$ and $\Delta_{0}$, we
then obtain Eq.~(3.11) in Ref.~\citep{Mattis1958}:
\begin{align}
\widetilde{\sigma}_{1}^{\textrm{BCS}}\left(\nu;T=0\right) &= \Theta\left(\nu-2\Delta_{0}\right)\biggl[\left(1+\frac{2\Delta_{0}}{\nu}\right)E\left(k\right) \nonumber\\
&\quad -\frac{4\Delta_{0}}{\nu}K\left(k\right)\biggr],\quad k=\left|\frac{\nu-2\Delta_{0}}{\nu+2\Delta_{0}}\right|.
\label{eq:MBT0}
\end{align}
For frequencies $\nu<2\Delta_{0}$ the real-part of the conductivity vanishes at zero temperature. Mattis and Bardeen also give the result for the imaginary part of the conductivity in the $T=0$ limit.

\subsubsection{Normal-state limit}

In the normal state, $\Delta=0$ and so the expression in Eq.~\eqref{eq:ZimmEq} becomes
\begin{equation}
\sigma_{n}\left(\nu\right)=\sigma_{0}\frac{i}{2\nu\tau}\left(J+\int_{0}^{\infty}I_{2}d\omega\right),
\end{equation}
where 
\begin{equation}
J\left(\nu\geq0\right)=\int_{0}^{\nu}I_{3}d\omega.
\end{equation}
In the normal state $P_{1}=\omega+\nu,P_{2}=\omega$, and $P_{3}=\nu-\omega$.
The integral over $I_{2}$ is given by 
\begin{align}
\int_{0}^{\infty}I_{2}d\omega &= \lim_{L\rightarrow\infty}\frac{2}{\nu+i/\tau}\int_{0}^{L}\biggl[\tanh\left(\frac{\omega+\nu}{2T}\right) \nonumber\\
&\quad -\tanh\left(\frac{\omega}{2T}\right)\biggr]d\omega\nonumber \\
 &=  \lim_{L\rightarrow\infty}\frac{2}{\nu+i/\tau}\left(2T\right)\biggl[\log\cosh\left(\frac{L+\nu}{2T}\right) \nonumber\\
 &\quad -\log\cosh\left(\frac{\nu}{2T}\right)-\log\cosh\left(\frac{L}{2T}\right)\biggr].
\end{align}
The integral over $I_{3}$ is given by 
\begin{align}
\int_{0}^{\nu}I_{3}d\omega &= \frac{2}{\nu+i/\tau}\int_{0}^{\nu}\tanh\left(\frac{\omega}{2T}\right)d\omega\nonumber \\
 &= \frac{2}{\nu+i/\tau}\left(2T\right)\log\cosh\left(\frac{\nu}{2T}\right).
\end{align}
The normal-state conductivity is thus 
\begin{align}
\sigma_{n}\left(\nu\right) &= \sigma_{0}\frac{i}{2\nu\tau}\lim_{L\rightarrow\infty}\frac{2}{\nu+i/\tau}\left(2T\right)\biggl[\log\cosh\left(\frac{L+\nu}{2T}\right) \nonumber\\
&\quad -\log\cosh\left(\frac{L}{2T}\right)\biggr]\nonumber \\
 &= \sigma_{0}\frac{i}{2\nu\tau}\frac{2}{\nu+i/\tau}2T\frac{\nu}{2T}\nonumber \\
 &= \sigma_{0}\left(\frac{1/\tau^{2}}{\nu^{2}+1/\tau^{2}}+\frac{i\nu/\tau}{\nu^{2}+1/\tau^{2}}\right).
\end{align}
The DC normal-state conductivity is given by $\sigma_{n}\left(\nu=0\right)\equiv\sigma_{0}= ne^{2}\tau/m$.
Thus, the real part of the normal-state conductivity is given by 
\begin{equation}
\textrm{Re}\left[\sigma_{n}\left(\nu\right)\right]=\frac{\sigma_{0}}{1+\left(\nu\tau\right)^{2}}.
\end{equation}

\subsubsection{Additional numerical results for the BCS electrical conductivity}

\begin{figure*}[t]
\centering\includegraphics[width=13.5cm,height=13.5cm]{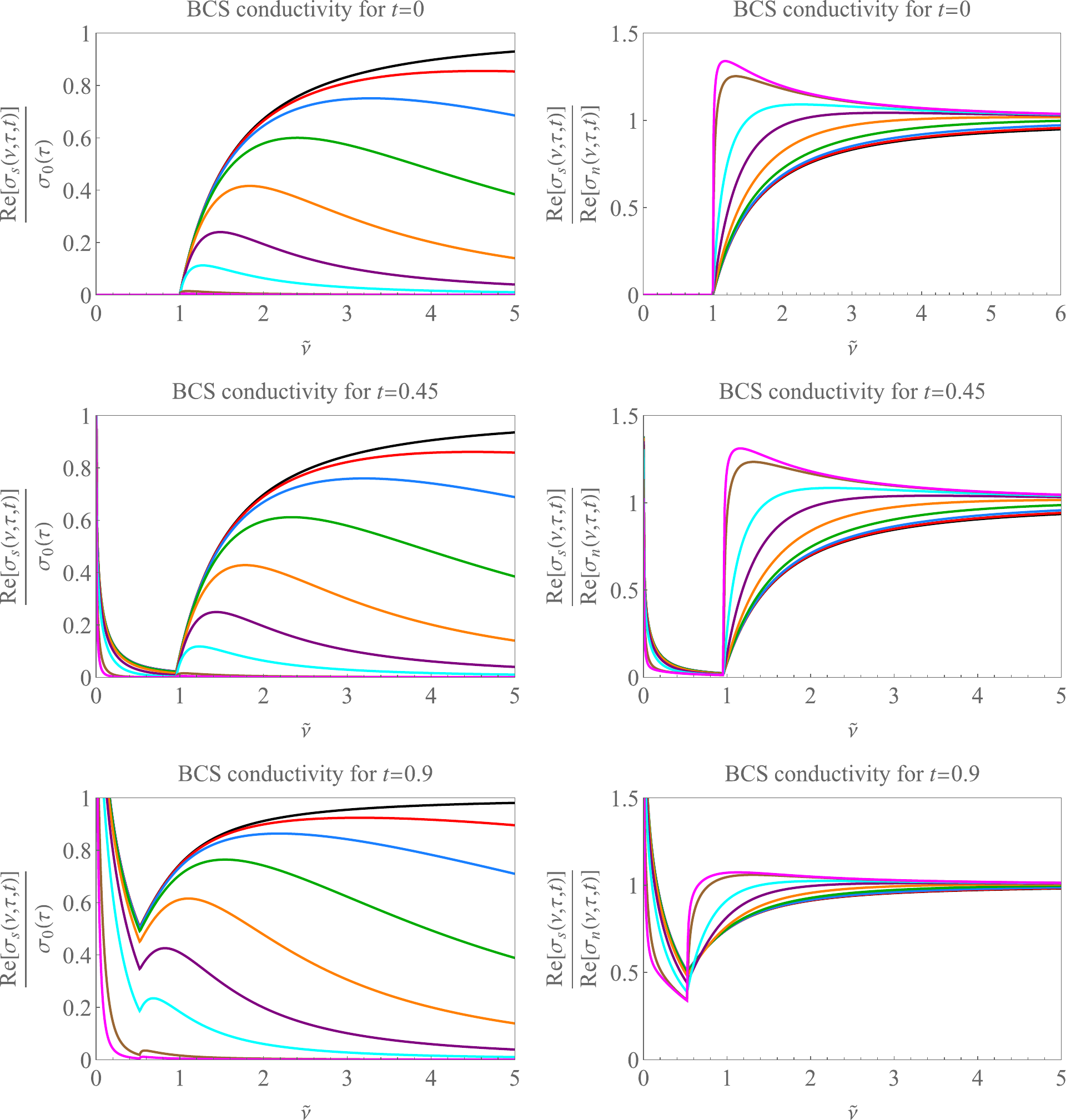}
\caption{Zero-temperature and finite reduced temperature plots ($t\equiv T/T_{c}=0.45$ and $t=0.9$) of the real part of the BCS electrical conductivity. Here we use units defined by $\widetilde{\nu}\equiv\nu/\left(2\Delta_{0}\right)$ and $y=1/\left(2\tau\Delta_{0}\right)$, where $\Delta_{0}$ is the zero-temperature BCS gap. The $y$ values (in order from the top curve (black) to the bottom curve (fuchsia) are given by $y=500$ (dirty limit)$, 16, 8, 4, 2, 1, 1/2, 1/4, 1/8$ (clean limit). In the figures on the left, we normalize by $\sigma_{0}\equiv ne^2\tau/m$, whereas in the figures on the right we normalize by the normal-state results (which have $\Delta=0$).}
\label{fig:BCSPlots1}
\end{figure*}

In this section we provide additional plots of the BCS electrical conductivity as a function of the external frequency $\nu$. We define a dimensionless frequency $\widetilde{\nu}\equiv \nu/\left(2\Delta_{0}\right)$ and a dimensionless impurity parameter $y=1/\left(2\tau\Delta_{0}\right)$. In Fig.~\ref{fig:BCSPlots1} we plot $\textrm{Re}\left[\sigma_{s}\left(\nu\right)\right]$, both in units of $\sigma_{0}$ and $\textrm{Re}\left[\sigma_{n}\left(\nu\right)\right]$. We show plots of the normalized conductivity as a function of $\widetilde{\nu}$ for various $y$ values, at reduced temperatures $t=0, t=0.45,$ and $t=0.9$. 

\clearpage

\bibliographystyle{naturemag}

\clearpage


\section*{Supplementary material}

# Supplementary Information: Electrical conductivity and Nuclear magnetic resonance relaxation rate of Eliashberg superconductors in the weak-coupling limit

Rufus Boyack,[1] Sepideh Mirabi,[2] and F. Marsiglio[3]

[1]Department of Physics and Astronomy, Dartmouth College, Hanover, New Hampshire, 03755, USA
[2]Department of Physics, Simon Fraser University, Burnaby, British Columbia, V5A 1S6, Canada
[3]Department of Physics, University of Alberta, Edmonton, Alberta, T6G 2E1, Canada

## SUPPLEMENTARY NOTE 1

In this subsection, we present more details relating to the analysis of $\Delta_2$. We consider an Einstein model [1] with electron-phonon coupling $\alpha^2 F(\omega) = A\delta(\omega - \omega_E)$, where $A = \lambda \omega_E/2$, and the bosonic propagator is $\lambda(z) = 2A\omega_E/(\omega_E^2 - z^2)$. The real frequency axis Eliashberg equations are then given by [1]:

$$\Delta(\omega + i\delta)Z(\omega + i\delta) = \pi T \sum_{m=-\infty}^{\infty} \lambda(\omega - i\omega_m) \frac{\Delta(i\omega_m)}{\sqrt{\omega_m^2 + \Delta^2(i\omega_m)}}$$
$$+ \frac{i\pi}{2}\lambda \omega_E \Bigg\{ [N(\omega_E) + f(\omega_E - \omega)] \frac{\Delta(\omega - \omega_E + i\delta)}{\sqrt{(\omega - \omega_E + i\delta)^2 - \Delta^2(\omega - \omega_E + i\delta)}}$$
$$+ [N(\omega_E) + f(\omega_E + \omega)] \frac{\Delta(\omega + \omega_E + i\delta)}{\sqrt{(\omega + \omega_E + i\delta)^2 - \Delta^2(\omega + \omega_E + i\delta)}} \Bigg\}, \quad (0.1)$$

$$Z(\omega + i\delta) = 1 + \frac{i\pi T}{\omega} \sum_{m=-\infty}^{\infty} \lambda(\omega - i\omega_m) \frac{\omega_m}{\sqrt{\omega_m^2 + \Delta^2(i\omega_m)}}$$
$$+ i\pi\lambda \frac{\omega_E}{2\omega} \Bigg\{ [N(\omega_E) + f(\omega_E - \omega)] \frac{(\omega - \omega_E)}{\sqrt{(\omega - \omega_E + i\delta)^2 - \Delta^2(\omega - \omega_E + i\delta)}}$$
$$+ [N(\omega_E) + f(\omega_E + \omega)] \frac{(\omega + \omega_E)}{\sqrt{(\omega + \omega_E + i\delta)^2 - \Delta^2(\omega + \omega_E + i\delta)}} \Bigg\}. \quad (0.2)$$

Using the definition of $\lambda(z)$, and the fact that $\Delta$ is purely real on the imaginary frequency axis, one can show that the summation term in Eq. (0.1) is purely real and the summation term in Eq. (0.2) is purely imaginary (without including the prefactor $i$ that is).

We define $\Delta(\omega = \Delta_1, T) = \Delta_1(T) + i\Delta_2(T)$ and $Z(\omega = \Delta_1, T) = Z_1(T) + iZ_2(T)$. Insert these definitions into Eqs. (0.1)-(0.2) and then take the imaginary part of both sides of the equations. As mentioned earlier, the first summation is purely real and the second summation is purely imaginary; thus, we obtain:

$$Z_1\Delta_2 + Z_2\Delta_1 = \pi\lambda\omega_E \frac{1}{2} \text{Re}\Bigg\{ [N(\omega_E) + f(\omega_E - \Delta_1)] \frac{\Delta(\Delta_1 - \omega_E + i\delta)}{\sqrt{(\Delta_1 - \omega_E + i\delta)^2 - \Delta^2(\Delta_1 - \omega_E + i\delta)}}$$
$$+ [N(\omega_E) + f(\omega_E + \Delta_1)] \frac{\Delta(\Delta_1 + \omega_E + i\delta)}{\sqrt{(\Delta_1 + \omega_E + i\delta)^2 - \Delta^2(\Delta_1 + \omega_E + i\delta)}} \Bigg\}, \quad (0.3)$$

$$Z_2 = \pi\lambda \frac{\omega_E}{2\Delta_1} \text{Re}\Bigg\{ [N(\omega_E) + f(\omega_E - \Delta_1)] \frac{(\Delta_1 - \omega_E)}{\sqrt{(\Delta_1 - \omega_E + i\delta)^2 - \Delta^2(\Delta_1 - \omega_E + i\delta)}}$$
$$+ [N(\omega_E) + f(\omega_E + \Delta_1)] \frac{(\Delta_1 + \omega_E)}{\sqrt{(\Delta_1 + \omega_E + i\delta)^2 - \Delta^2(\Delta_1 + \omega_E + i\delta)}} \Bigg\}. \quad (0.4)$$

The next step is to use Eq. (0.4) to eliminate $Z_2$ from Eq. (0.3); doing this results in:

$$Z_1\Delta_2 = \pi\lambda\omega_E \frac{1}{2}\text{Re}\left\{[N(\omega_E) + f(\omega_E - \Delta_1)]\frac{\omega_E - \Delta_1 + \Delta(\Delta_1 - \omega_E + i\delta)}{\sqrt{(\Delta_1 - \omega_E + i\delta)^2 - \Delta^2(\Delta_1 - \omega_E + i\delta)}}\right.$$
$$\left. + [N(\omega_E) + f(\omega_E + \Delta_1)]\frac{\Delta(\Delta_1 + \omega_E + i\delta) - (\Delta_1 + \omega_E)}{\sqrt{(\Delta_1 + \omega_E + i\delta)^2 - \Delta^2(\Delta_1 + \omega_E + i\delta)}}\right\}. \quad (0.5)$$

Finally, the sign of the first square root is chosen to be the same as the sign of $\Delta_1 - \omega_E < 0$; thus, the first square root has a negative sign. Accounting for this fact, we can write Eq. (0.5) as

$$Z_1\Delta_2 = -\pi\lambda\omega_E \frac{1}{2}\text{Re}\left\{[N(\omega_E) + f(\omega_E - \Delta_1)]\frac{\omega_E - \Delta_1 + \Delta(\omega_E - \Delta_1 + i\delta)}{\sqrt{(\omega_E - \Delta_1 + i\delta)^2 - \Delta^2(\omega_E - \Delta_1 + i\delta)}}\right.$$
$$\left. + [N(\omega_E) + f(\omega_E + \Delta_1)]\frac{\Delta_1 + \omega_E - \Delta(\Delta_1 + \omega_E + i\delta)}{\sqrt{(\Delta_1 + \omega_E + i\delta)^2 - \Delta^2(\Delta_1 + \omega_E + i\delta)}}\right\}. \quad (0.6)$$

The square roots in Eq. (0.6) have positive signs. Rearranging Eq. (0.6) then gives Eq. (2.22) in the main text.

To write $\Delta_2$ explicitly in terms of the parameters of the theory ($\omega_E$ and $\lambda$) we must determine $\Delta(\omega_E \pm \Delta_1)$. In Ref. [2] we considered an approximation:

$$\Delta_{\text{approx}}(\omega + i\delta) = \frac{\Delta_1}{1 - \overline{\omega}^2}[1 + \lambda h(\omega)], \quad (0.7)$$

where $h(\omega)$ is given by [2]:

$$h(\omega) = \frac{3}{2} - \frac{1}{4\overline{\omega}}\text{Re}\left[\psi\left(\frac{1}{2} + i\left(\frac{\omega_E + \omega}{2\pi T}\right)\right) - \psi\left(\frac{1}{2} + i\left(\frac{\omega_E - \omega}{2\pi T}\right)\right)\right]$$
$$- \frac{3}{4}\frac{1}{2 - \overline{\omega}}\text{Re}\left[\psi\left(\frac{1}{2} + i\frac{\omega_E}{2\pi T}\right) - \psi\left(\frac{1}{2} + i\left(\frac{\omega_E - \omega}{2\pi T}\right)\right)\right]$$
$$- \frac{3}{4}\frac{1}{2 + \overline{\omega}}\text{Re}\left[\psi\left(\frac{1}{2} + i\frac{\omega_E}{2\pi T}\right) - \psi\left(\frac{1}{2} + i\left(\frac{\omega_E + \omega}{2\pi T}\right)\right)\right], \quad \overline{\omega} \equiv \omega/\omega_E. \quad (0.8)$$

The expression in Eq. (0.7) is obtained by splitting up the interaction $\lambda(\omega - i\omega_m)$ in Eq. (0.1) into two different parts and performing the frequency summation using digamma functions; see Refs. [2, 3] for more details. In the immediate vicinity of $\omega \approx \omega_E$, Eq. (0.7) is not valid. This is because in deriving Eq. (0.7) we have omitted the terms in Eq. (0.1) that depend on $\Delta(\omega \pm \omega_E)$. In Supplementary Table I, we present some numerical values of $\Delta(\omega \pm \Delta_1)$ to illustrate how rapidly $\Delta$ changes for frequencies near $\omega_E$. While at $\omega = \omega_E$ the gap diverges, at $\omega = \omega_E \pm \Delta_1$ the gap is $\mathcal{O}(1)$, as indicated in the table. In Supplementary Table II, we compute $\Delta(\omega_E \pm \Delta_1)$ based on Eq. (0.7), which should be compared with the numerical results listed in Supplementary Table I. The approximation in Eq. (0.7) is clearly inapplicable. Determining $\Delta_2$ analytically, for arbitrary temperatures, remains a problem for future work. Let us investigate a possible solution for $t \to 1$.

As shown in Fig. 7 of Ref. [2], for fixed frequency, as $T/T_c \to 1$, the order parameter $\Delta$ goes to zero. Assuming this is the case here, in Eq. (2.22) of the main text we can drop the $\Delta(\omega_E \pm \Delta_1)$ terms to obtain:

$$(1 + \lambda)\frac{\Delta_2}{\omega_E} \approx -2\pi\lambda\exp\left(-\frac{\omega_E}{T_c t}\right), \quad \frac{T}{T_c} - 1 \ll 1. \quad (0.9)$$

As a numerical check of this expression, when $\lambda = 1$ and $t = 0.99$, the numerical value of $\Delta_2$ is $\Delta_2 \approx -9.7 \times 10^{-5}\omega_E$ whereas the above formula gives $\Delta_2 \approx -4.7 \times 10^{-4}\omega_E$. While there is a noticeable discrepancy between these values, we have found that if the electron-phonon spectrum is altered from a Dirac delta function to a Lorentzian, then the numerical value for $\Delta_2$ at $t = 0.99$ for $\lambda = 1$ is $-4.4 \times 10^{-4}\omega_E$, which is quite close to the theoretical prediction above. We leave it to future work to further investigate the behaviour of $\Delta_2$ for a Lorentzian spectrum. Bandwidth effects on conductivity and NMR have been investigated respectively in Ref. [4] and Ref. [5].

|  | $\Delta(\omega)$ | $Z(\omega)$ | $\Delta(\omega_E - \omega)$ | $\Delta(\omega_E + \omega)$ | $\Delta_2$ |
|---|---|---|---|---|---|
| $t=0.98$; $\overline{\omega} = 0.06185$ | $0.06185$ $-7.74439 \times 10^{-5} i$ | $2.05592$ $+1.74270 \times 10^{-2} i$ | $-1.07427$ $-1.88476 \times 10^{-1} i$ | $0.96720$ $+7.72750 \times 10^{-2} i$ | $-7.74278 \times 10^{-5}$ |
| $t=0.96$; $\overline{\omega} = 0.08667$ | $0.08665$ $-6.66516 \times 10^{-5} i$ | $2.04688$ $+1.19539 \times 10^{-2} i$ | $-1.00818$ $-1.77077 \times 10^{-1} i$ | $1.02248$ $+7.06923 \times 10^{-2} i$ | $-6.67474 \times 10^{-5}$ |
| $t=0.94$; $\overline{\omega} = 0.10515$ | $0.10515$ $-5.03438 \times 10^{-5} i$ | $2.04022$ $+8.87636 \times 10^{-3} i$ | $-0.99580$ $-1.41406 \times 10^{-1} i$ | $1.05275$ $+5.01485 \times 10^{-2} i$ | $-5.03569 \times 10^{-5}$ |
| $t=0.92$; $\overline{\omega} = 0.12028$ | $0.12027$ $-4.51663 \times 10^{-5} i$ | $2.03305$ $+6.96118 \times 10^{-3} i$ | $-0.96900$ $-1.43928 \times 10^{-1} i$ | $1.07726$ $+4.69440 \times 10^{-2} i$ | $-4.51936 \times 10^{-5}$ |
| $t=0.90$; $\overline{\omega} = 0.13318$ | $0.13318$ $-3.43911 \times 10^{-5} i$ | $2.02698$ $+5.43201 \times 10^{-3} i$ | $-0.96341$ $-1.19837 \times 10^{-1} i$ | $1.09522$ $+3.48618 \times 10^{-2} i$ | $-3.43841 \times 10^{-5}$ |

Supplementary Table I: Numerically determined gap ($\Delta$) and renormalization ($Z$) functions. The electron-phonon coupling is $\lambda = 1$, $T_c = 0.11462\omega_E$. The frequencies and the gap are in units of $\omega_E$. The frequencies used are close to the gap edge. The final column shows the value of $\Delta_2$ as computed using Eq. (2.22), based on the numerical values of $\Delta(\omega_E \pm \Delta_1)$ given in the table; the values agree with the imaginary part in the second column within numerical uncertainty.

|  |  | $\Delta_{\text{approx}}(\omega_E - \omega)$ | $\Delta_{\text{approx}}(\omega_E + \omega)$ |
|---|---|---|---|
| $t=0.98$; | $\overline{\omega} = 0.06185$ | $0.168738$ | $-0.0987065$ |
| $t=0.96$; | $\overline{\omega} = 0.08667$ | $0.199892$ | $-0.101679$ |
| $t=0.94$; | $\overline{\omega} = 0.10515$ | $0.227103$ | $-0.107856$ |
| $t=0.92$; | $\overline{\omega} = 0.12028$ | $0.252103$ | $-0.115595$ |
| $t=0.90$; | $\overline{\omega} = 0.13318$ | $0.275468$ | $-0.124210$ |

Supplementary Table II: The approximate gap computed via Eqs. (0.7)-(0.8) for $\lambda = 1$, $T_c = 0.11462\omega_E$. Comparing these values with those in Supplementary Table I shows that, near $\omega = \omega_E \pm \Delta_1$, the approximation (0.7) is invalid.

### SUPPLEMENTARY REFERENCES


\clearpage

\end{document}